\documentclass[12pt]{article}
\usepackage{cite}
\usepackage{accents}

\renewcommand{\theequation}
{\arabic{section}.\arabic{equation}}

\makeatletter
\def\eqnarray{ \stepcounter{equation} \let\@currentlabel=\theequation
 \global\@eqnswtrue
 \global\@eqcnt\z@
 \tabskip\@centering
 \let\\=\@eqncr
 $$\halign to \displaywidth\bgroup\@eqnsel\hskip\@centering
 $\displaystyle\tabskip\z@{##}$&\global\@eqcnt\@ne
 \hfil$\displaystyle{{}##{}}$\hfil
 &\global\@eqcnt\tw@$\displaystyle\tabskip\z@{##}$\hfil
 \tabskip\@centering&\llap{##}\tabskip\z@\cr}
\makeatother

\makeatletter
\def\@arrayacol{\edef\@preamble{\@preamble \hskip .5\arraycolsep}}
\def\array{\let\@acol\@arrayacol \let\@classz\@arrayclassz
\let\@classiv\@arrayclassiv \let\\\@arraycr\def\@halignto{}\@tabarray}
\makeatother

\renewcommand{\arraystretch}{1.6}

\makeatletter
\newcounter{subeqncnt}
\def\thesubeqncnt{\alph{subeqncnt}}
\def\subequations{\begingroup%
   \stepcounter{equation}\edef\@tempa{\theequation}%
   \let\c@equation\c@subeqncnt\c@subeqncnt\z@
   \edef\theequation{\@tempa\noexpand\thesubeqncnt}}

\makeatother

\newcommand{\be}{\begin{equation}}
\newcommand{\ee}{\end{equation}}

\newcommand{\beqa}{\begin{eqnarray}}
\newcommand{\eeqa}{\end{eqnarray}}
\newcommand{\nn}{\nonumber}

\newcommand{\eqref}[1]{(\ref{#1})}

\def\CL {{\cal L}}

\newcommand{\del}{\partial}

\newcommand{\tr}{{\rm Tr}\,}

\begin{document}

\setlength{\baselineskip}{7mm}
\begin{titlepage}
\begin{flushright}
{\tt NRCPS-HE-05-63} \\
December, 2005
\end{flushright}

\vspace{1cm}

\begin{center}
{\Large Gauge Invariant Lagrangian \\
for \\
Non-Abelian Tensor Gauge Fields of Fourth Rank
}

\vspace{1cm}

{\sc{George Savvidy\footnote{savvidy(AT)inp.demokritos.gr}}}
and
{\sc{Takuya Tsukioka\footnote{tsukioka(AT)inp.demokritos.gr}}}
\\
{\it Institute of Nuclear Physics,} \\
{\it National Center for Scientific Research,}
{\it ``DEMOKRITOS''}, \\
{\it Agia Paraskevi, GR-15310 Athens, Greece} 
\end{center}

\vspace{1cm}

\begin{abstract}
Using generalized field strength tensors for non-Abelian tensor gauge fields
one can explicitly construct all possible Lorentz invariant
quadratic forms for rank-4 non-Abelian tensor gauge fields and
demonstrate that there exist only two linear combinations of them which form
a gauge invariant Lagrangian. Together with the previous construction of
independent gauge invariant forms for rank-2 and rank-3 tensor gauge fields this
construction proves the uniqueness of early proposed general Lagrangian up to
rank-4 tensor fields. Expression for the coefficients
of the general Lagrangian is presented in a compact form.
\end{abstract}

\end{titlepage}

\section{Introduction}
\setcounter{equation}{0}
\setcounter{footnote}{0}

The early investigation of higher-spin representations of the Poincar\'e
algebra and of the corresponding field equations is due to Majorana,
Dirac, Fierz and Pauli and Wigner \cite{majorana,dirac,fierzpauli,wigner}.
The theory of massive particles
of higher spin was developed by Fierz and Pauli \cite{fierzpauli} and
Rarita and Schwinger \cite{rarita}. The Lagrangian and
S-matrix formulations of {\it free field
theory} of massive and massless fields with higher spin
have been completely constructed in
\cite{yukawa1,schwinger,Weinberg:1964cn,Weinberg:1964ev,Weinberg:1964ew,
chang,singh,singh1,fronsdal,fronsdal1}.
The problem of {\it introducing interaction} appears to be much more complex
\cite{Gupta,kraichnan,thirring,feynman,deser,fronsdal2,Sezgin:2001zs,Sagnotti:2005ns}
and met enormous difficulties for spin fields higher than two
\cite{witten,deser1,berends,dewit,vasiliev}.
The first positive result in this direction was
the light-front construction of the cubic
interaction term for the massless field of helicity $\pm \lambda$ in
\cite{Bengtsson:1983pd,Bengtsson:1983pg}.
Discussion of the tensionless strings and their relation with higher spin
fields can be found in \cite{Edgren:2005gq,Turok:2004gb,Engquist:2005yt,
Bekaert:2005vh,Brink:2005wh,Gamboa:2004cv,Bonelli:2004ve,Bakas:2004jq,
Bredthauer:2004kv,Gamboa:2003fy,Chagas-Filho:2003wv,Bianchi:2003wx}.
Alternative formulations of higher spin field theories have been proposed and
discussed in \cite{Gabrielli:1990ay,Kawamoto:1990hi,Gabrielli:1999xt,Baez:2002jn,
Ivanov:1976pg,Ivanov:1979ny,Curtright:1987zc,Manvelyan:2005ew}.

In the recent articles \cite{Savvidy:2005zm, Savvidy:2005fi, Savvidy:2005ki}
one of the authors considered the generalization of the Yang-Mills theory
which includes non-Abelian tensor gauge fields of higher rank. For
non-Abelian tensor gauge fields of arbitrary higher rank-s
two independent gauge invariant forms $\CL_{s}$ and $\CL_{s}'$
which are quadratic in field strength tensors have been found.
The general Lagrangian ${{\cal L}}$ was defined as a linear sum
of these forms \cite{Savvidy:2005fi, Savvidy:2005ki}
$$
{{\cal L}} = \sum^{\infty}_{s=0}~ g_{s+1} {{\cal L}}_{s+1}~+
\sum^{\infty}_{s=1}~ g^{'}_{s+1} {{\cal L}}^{'}_{s+1}.
$$
In order to prove a uniqueness of the general Lagrangian it is important to know whether
these invariants provide {\it a complete set of independent
gauge invariants} quadratic in field strength tensors. The
affirmative answer to the above question will prove that it is
indeed a unique gauge invariant Lagrangian.

In this article, using generalized field strength tensors for
non-Abelian tensor gauge fields defined in
\cite{Savvidy:2005zm, Savvidy:2005fi, Savvidy:2005ki},
we explicitly constructed all possible Lorentz invariant
quadratic forms for the rank-4 non-Abelian tensor gauge fields and
demonstrated that there are only two linear combinations of
them ${{\cal L}}_{4}$ and ${{\cal L}}^{'}_{4}$ which form a gauge invariant Lagrangian.
Together with the previous construction of independent
gauge invariant forms for rank-2 (${{\cal L}}_{2}$ and ${{\cal L}}^{'}_{2}$)
and rank-3 (${{\cal L}}_{3}$ and ${{\cal L}}^{'}_{3}$) tensor gauge fields
\cite{Savvidy:2005fi, Savvidy:2005ki}, the above consideration
extends the proof of the uniqueness of the total
Lagrangian up to the rank-4 gauge fields.

It seems difficult to extend this explicit construction of all possible Lorentz invariant
structures to higher rank tensor fields because their number grows very fast.
Nevertheless the analysis of low rank tensor gauge fields makes it
plausible that the general Lagrangian presented in
\cite{Savvidy:2005fi, Savvidy:2005ki} is indeed a unique one
and contains only two independent gauge invariant forms $\CL_{s}$ and $\CL_{s}'$.

This paper is organized as follows: in the next section
necessary notations and definitions of extended gauge transformation
of tensor gauge fields and of the corresponding field strength
tensors from the articles \cite{Savvidy:2005zm, Savvidy:2005fi, Savvidy:2005ki}
will be introduced. The previous construction of independent gauge invariant forms
for the rank-2 and rank-3 tensor fields will be reproduced in Section 3.
In the subsequent, main part of the article,
we shall construct all Lorentz invariant
quadratic forms for the rank-4 tensor fields and
shall prove that only two forms
$\CL_4$ and $\CL_4'$ provide all possible invariants
\begin{eqnarray*}
\CL_4
&=&
\tr
\bigg(
G_{\mu\nu, \rho\sigma\lambda}G_{\mu\nu, \rho\sigma\lambda}
+\frac{3}{2}
G_{\mu\nu, \rho\rho\sigma}G_{\mu\nu, \sigma\lambda\lambda}
+3
G_{\mu\nu, \rho\sigma}G_{\mu\nu, \rho\sigma\lambda\lambda}
\nonumber
\\
&&\hspace*{8mm}+\frac{3}{4}
G_{\mu\nu, \rho\rho}G_{\mu\nu, \sigma\sigma\lambda\lambda}
+\frac{3}{2}
G_{\mu\nu, \rho}G_{\mu\nu, \rho\sigma\sigma\lambda\lambda}
+\frac{1}{4}
G_{\mu\nu}G_{\mu\nu, \rho\rho\sigma\sigma\lambda\lambda}
\bigg),
\end{eqnarray*}
and
\begin{eqnarray*}
\CL_4'
&=&
\tr
\bigg(
G_{\mu\nu, \rho\sigma\lambda}G_{\mu\rho, \nu\sigma\lambda}
+\frac{1}{2}
G_{\mu\nu, \rho\sigma\sigma}G_{\mu\rho, \nu\lambda\lambda}
+2
G_{\mu\nu, \rho\sigma\sigma}G_{\mu\lambda, \nu\rho\lambda}
+
G_{\mu\nu, \nu\rho\sigma}G_{\mu\lambda, \rho\sigma\lambda}
\\
&&
\hspace*{8mm}
+\frac{1}{2}
G_{\mu\nu, \nu\rho\rho}G_{\mu\sigma, \sigma\lambda\lambda}
+2
G_{\mu\nu, \rho\sigma}G_{\mu\rho, \nu\sigma\lambda\lambda}
+2
G_{\mu\nu, \rho\sigma}G_{\mu\lambda, \nu\rho\sigma\lambda}
+
G_{\mu\nu, \rho\rho}G_{\mu\sigma, \nu\sigma\lambda\lambda}
\nonumber
\\
&&
\hspace*{8mm}
+\frac{1}{2}
G_{\mu\nu, \nu\rho}G_{\mu\rho, \lambda\lambda\sigma\sigma}
+2
G_{\mu\nu, \nu\rho}G_{\mu\lambda, \rho\lambda\sigma\sigma}
+\frac{1}{2}
G_{\mu\nu, \rho}G_{\mu\rho, \nu\lambda\lambda\sigma\sigma}
+2
G_{\mu\nu, \rho}G_{\mu\lambda, \nu\rho\lambda\sigma\sigma}
\nn\\
&&
\hspace*{8mm}
+\frac{1}{2}
G_{\mu\nu, \nu}G_{\mu\rho, \rho\sigma\sigma\lambda\lambda}
\nonumber
+\frac{1}{2}
G_{\mu\nu}G_{\mu\rho, \nu\rho\sigma\sigma\lambda\lambda}
\bigg).
\end{eqnarray*}
This consideration allows to define a unique gauge
invariant Lagrangian for the fourth-rank tensor gauge fields which is a
linear sum of $\CL_4$ and $\CL_4'$.

In Section 5 we shall demonstrate that in the general Lagrangian
\cite{Savvidy:2005fi, Savvidy:2005ki} the total number of Lorentz independent structures
for the rank-$s$ tensor gauge field grows as $s^2$ and
shall present a compact expression for the coefficients.

\section{Gauge Fields and Field Strengths}
\setcounter{equation}{0}
\setcounter{footnote}{0}

The Lie algebra valued non-Abelian gauge fields
are defined as rank-($s+1$) tensors~\cite{Savvidy:2005zm, Savvidy:2005fi, Savvidy:2005ki}
$$
A_{\mu\lambda_1\cdots\lambda_s}
=A^a_{\mu\lambda_1\cdots\lambda_s}L_a,
\qquad
s=0, 1, 2, \cdots,
$$
where $L_a$ are generators of the compact Lie gauge group $G$.
These fields are totally symmetric with respect to the
indices $\lambda_1\cdots\lambda_{s}$.
{\it A priory}, the tensor gauge fields
have no symmetric and/or antisymmetric structures with
respect to the first index $\mu$.

The extended gauge transformations of the non-Abelian tensor gauge
fields are defined by the following equations \cite{Savvidy:2005zm}
\begin{subequations}
\begin{equation}
\begin{array}{rcl}
\delta A_\mu
&=&
\del_\mu\xi
-ig
[A_\mu, \ \xi],
\\
\delta A_{\mu\lambda_1}
&=&
\del_\mu\xi_{\lambda_1}
-ig
[A_\mu, \ \xi_{\lambda_1}]
-ig
[A_{\mu\lambda_1}, \ \xi],
\\
\delta A_{\mu\lambda_1\lambda_2}
&=&
\del_\mu\xi_{\lambda_1\lambda_2}
-ig
[A_\mu, \ \xi_{\lambda_1\lambda_2}]
-ig
[A_{\mu\lambda_1}, \ \xi_{\lambda_2}]
-ig
[A_{\mu\lambda_2}, \ \xi_{\lambda_1}]
-ig
[A_{\mu\lambda_1\lambda_2}, \ \xi],
\\
\cdots
&\cdot& \cdots
\end{array}
\label{polygauge_0}
\end{equation}
or in the general form by the formula
\begin{equation}
\delta A_{\mu\lambda_1\cdots\lambda_s}
=
\del_\mu\xi_{\lambda_1\cdots\lambda_s}
-ig\sum_{i=0}^s
\sum_{p_1<p_2<\cdots<p_i}\!\!
\Big[
A_{\mu\lambda_{p_1}\lambda_{p_2}\cdots\lambda_{p_i}}, \
\xi_{\lambda_1
\cdots\accentset{\vee}{\lambda_{p_1}}
\cdots\accentset{\vee}{\lambda_{p_2}}
\cdots\accentset{\vee}{\lambda_{p_i}}
\cdots\lambda_s}
\Big],
\label{polygauge_1}
\end{equation}
\end{subequations}

\vspace*{-5mm}

\noindent
where the symbol ``$\accentset{\vee}{\ \ }$'' in
the $\accentset{\vee}{\lambda_{p_i}}$ means
that one should erase the index $\lambda_{p_i}$.
The infinitesimal non-Abelian gauge
parameters $\xi_{\lambda_{1}\cdots\lambda_s}(x)$ are
totally symmetric rank-$s$ tensors.
By construction, there are $\displaystyle\frac{s!}{i!(s-i)!}$ terms
in the sum $\displaystyle\sum_{p_1<p_2<\cdots<p_i}$.
These extended gauge transformations generate a closed algebraic
structure \cite{Savvidy:2005zm}
\begin{equation}
[\delta_{\xi_1}, \ \delta_{\xi_2}]A_{\mu\lambda_1\cdots\lambda_s}
=
\delta_{\xi}A_{\mu\lambda_1\cdots\lambda_s},
\label{gauge_algebra}
\end{equation}
where the gauge parameters $\xi_{\nu_1\cdots\nu_n}(x)$
on the right hand side are given by the formulas
\begin{subequations}
\begin{equation}
\begin{array}{rcl}
\xi
&=&
-ig
[\xi_1, \ \xi_2],
\\
\xi_{\nu_1}
&=&
-ig
[\xi_1, \ \xi_{2\nu_1}]
-ig
[\xi_{1\nu_1}, \xi_2],
\\
\xi_{\nu_1\nu_2}
&=&
-ig
[\xi_1, \ \xi_{2\nu_1\nu_2}]
-ig
[\xi_{1\nu_1}, \ \xi_{2\nu_2}]
-ig
[\xi_{1\nu_2}, \ \xi_{2\nu_1}]
-ig
[\xi_{1\nu_1\nu_2}, \ \xi_{2}],
\\
\cdots
&\cdot&
\cdots
\end{array}
\end{equation}
or in the general form by the formula
\begin{equation}
\xi_{\nu_1\cdots\nu_n}
=
-ig\sum_{i=0}^{n}\sum_{p_1<p_2<\cdots<p_i}\!\!
\Big[
{\xi_1}_{\nu_{p_1}\nu_{p_2}\cdots\nu_{p_i}}, \
{\xi_2}_{\nu_1\cdots
\accentset{\vee}{\nu_{p_1}}\cdots
\accentset{\vee}{\nu_{p_2}}\cdots
\accentset{\vee}{\nu_{p_i}}\cdots\nu_n}
\Big].
\end{equation}
\end{subequations}

\vspace*{-5mm}

The generalized field strengths are defined as \cite{Savvidy:2005zm}
\begin{subequations}
\begin{eqnarray}
G_{\mu\nu}
&=&
\del_\mu A_\nu
-\del_\nu A_\mu
-ig
[A_\mu, \ A_\nu],
\nonumber
\\
G_{\mu\nu, \lambda_1}
&=&
\del_\mu A_{\nu\lambda_1}
-\del_\nu A_{\mu\lambda_1}
-ig
[A_\mu, \ A_{\nu\lambda_1}]
-ig
[A_{\mu\lambda_1}, \ A_\nu],
\nonumber
\\
G_{\mu\nu, \lambda_1\lambda_2}
&=&
\del_\mu A_{\nu\lambda_1\lambda_2}
-\del_\nu A_{\mu\lambda_1\lambda_2}
\label{field_strength_0}
\\
&&
-ig
[A_\mu, \ A_{\nu\lambda_1\lambda_2}]
-ig
[A_{\mu\lambda_1}, \ A_{\nu\lambda_2}]
-ig
[A_{\mu\lambda_2}, \ A_{\nu\lambda_1}]
-ig
[A_{\mu\lambda_1\lambda_2}, \ A_\nu],
\nonumber
\\
\cdots
&\cdot&
\cdots
\nonumber
\end{eqnarray}
or in the general form by the formula
\begin{eqnarray}
G_{\mu\nu, \lambda_1\cdots\lambda_s}
&=&
\del_\mu A_{\nu\lambda_1\cdots\lambda_s}
-\del_\nu A_{\mu\lambda_1\cdots\lambda_s}
\nonumber
\\
&&
-ig\sum_{i=0}^{s}\sum_{p_1<p_2<\cdots<p_i}\!\!
\Big[
A_{\mu\lambda_{p_1}\lambda_{p_2}\cdots\lambda_{p_i}}, \
A_{\nu\lambda_1\cdots
\accentset{\vee}{\lambda_{p_1}}\cdots
\accentset{\vee}{\lambda_{p_2}}\cdots
\accentset{\vee}{\lambda_{p_i}}\cdots\lambda_s}
\Big].
\label{field_strength_1}
\end{eqnarray}
\end{subequations}

\vspace*{-5mm}

\noindent
These field strength tensors are antisymmetric in their first
two indices $(\mu, \nu)$
and are totally symmetric with respect to the rest of the
indices $(\lambda_1, \cdots, \lambda_s)$.
In the above definition of the extended gauge field strength
$G_{\mu\nu, \lambda_1\cdots\lambda_{s}}(x)$,
together with the classical Yang-Mills gauge boson
$A_{\mu}(x)$, there participate a set of higher rank gauge fields
$A_{\mu\lambda_1}(x)$, $A_{\mu\lambda_1\lambda_2}(x)$, $\cdots$,
$A_{\mu\lambda_1\cdots\lambda_{s}}(x)$ up to the rank $s+1$.
One of the important features of these field strengths
is that they transform homogeneously under the extended
gauge transformations (\ref{polygauge_0}) and (\ref{polygauge_1})
\cite{Savvidy:2005zm}:
\begin{subequations}
\begin{equation}
\begin{array}{rcl}
\delta G_{\mu\nu}
&=&
-ig
[G_{\mu\nu}, \ \xi],
\\
\delta G_{\mu\nu, \lambda_1}
&=&
-ig
[G_{\mu\nu}, \ \xi_{\lambda_1}]
-ig
[G_{\mu\nu, \lambda_1}, \ \xi],
\\
\delta G_{\mu\nu, \lambda_1\lambda_2}
&=&
-ig
[G_{\mu\nu}, \ \xi_{\lambda_1\lambda_2}]
-ig
[G_{\mu\nu, \lambda_1}, \ \xi_{\lambda_2}]
-ig
[G_{\mu\nu, \lambda_2}, \ \xi_{\lambda_1}]
-ig
[G_{\mu\nu, \lambda_1\lambda_2}, \ \xi],
\\
\cdots
&\cdot&
\cdots
\end{array}
\label{gauge_trans_field_strength_0}
\end{equation}
or in the general form:
\begin{equation}
\delta G_{\mu\nu, \lambda_1\cdots\lambda_s}
=
-ig\sum_{i=0}^{s}\sum_{p_1<p_2<\cdots<p_i}\!\!
\Big[
G_{\mu\nu, \lambda_{p_1}\lambda_{p_2}\cdots\lambda_{p_i}}, \
\xi_{\lambda_1\cdots
\accentset{\vee}{\lambda_{p_1}}\cdots
\accentset{\vee}{\lambda_{p_2}}\cdots
\accentset{\vee}{\lambda_{p_i}}\cdots\lambda_s}
\Big].
\label{gauge_trans_field_strength_1}
\end{equation}
\end{subequations}

\vspace*{-5mm}

The {\it gauge invariant tensor density} constructed in
\cite{Savvidy:2005fi, Savvidy:2005ki}
by expansion over the vector variable $e_{\mu}$ \cite{Savvidy:dv} is:
%
%
%
\begin{eqnarray}
\Big({\cal L}_{s+1}\Big)_{\nu_1\nu_2, \lambda_1\cdots\lambda_{2s}}
&=&
\tr
\Bigg(
\sum_{i=0}^{s-1}\sum_{p_1<p_2<\cdots<p_i}\!\!\!
G_{\mu\nu_1,
\lambda_{p_1}\lambda_{p_2}\cdots\lambda_{p_i}}
G_{\mu\nu_2,
\lambda_1\cdots\accentset{\vee}{\lambda_{p_1}}\cdots
\accentset{\vee}{\lambda_{p_2}}\cdots
\accentset{\vee}{\lambda_{p_i}}\cdots\lambda_{2s}}
\nonumber \\
&&
\hspace*{9mm}
+\frac{1}{2}\!\sum_{p_1<p_2<\cdots<p_s}\!\!\!
G_{\mu\nu_1, \lambda_{p_1}\lambda_{p_2}\cdots\lambda_{p_s}}
G_{\mu\nu_2,
\lambda_1\cdots\accentset{\vee}{\lambda_{p_1}}\cdots
\accentset{\vee}\lambda_{p_2}\cdots
\accentset{\vee}{\lambda_{p_s}}\cdots\lambda_{2s}}
\nonumber
\\
&&
\hspace*{9mm}
+(\nu_1\leftrightarrow\nu_2)
\Bigg).
\label{invariant_tensor_1}
\end{eqnarray}
These tensors are totally symmetric within the indices
$(\nu_1, \nu_2)$ and $(\lambda_1, \lambda_2, \cdots, \lambda_{2s})$,
whereas they have no symmetries between the
indices $\nu_i$ and $\lambda_j$.
Contraction of indices by metric tensors  allows to
construct two {\it gauge and also Lorentz invariant forms} ${\cal L}_{s+1}$
and ${\cal L}_{s+1}'$ \cite{Savvidy:2005fi, Savvidy:2005ki}
\begin{subequations}
\begin{eqnarray}
{\cal L}_{s+1}
&=&
\frac{1}{2^s}
\Big({\cal L}_{s+1}\Big)_{\nu_1\nu_2, \lambda_1\cdots\lambda_{2s}}
\Big(
\eta_{\nu_1\nu_2}\eta_{\lambda_1\lambda_2}
\Big)
\eta_{\lambda_3\lambda_4}
\cdots\eta_{\lambda_{2s-1}\lambda_{2s}},
\label{general_lagrangian_0}\\
{\cal L}_{s+1}'
&=&
\frac{1}{2^{s-1}}
\Big({\cal L}_{s+1}\Big)_{\nu_1\nu_2, \lambda_1\cdots\lambda_{2s}}
\Big(\eta_{\nu_1\lambda_1}\eta_{\nu_2\lambda_2}
\Big)
\eta_{\lambda_3\lambda_4}
\cdots\eta_{\lambda_{2s-1}\lambda_{2s}}.
\label{general_lagrangian_prime_0}
\end{eqnarray}
\end{subequations}
In order to define a unique Lagrangian it is important to know whether
 these invariants provide a {\it complete set of independent
gauge invariants} which are quadratic in field strength tensors. The
affirmative answer to the above question will prove that a unique
gauge invariant Lagrangian is \cite{Savvidy:2005fi, Savvidy:2005ki}
\be\label{fulllagrangian2}
{{\cal L}} = \sum^{\infty}_{s=0}~ g_{s+1} {{\cal L}}_{s+1}~+
\sum^{\infty}_{s=1}~ g^{'}_{s+1} {{\cal L}}^{'}_{s+1}~.
\ee

An alternative way of constructing a gauge invariant Lagrangian is, first, to
consider all possible  {\it Lorentz invariant quadratic forms},
and then by direct calculation of their variation over the gauge transformation
to check if some linear combination of them can be made gauge invariant
\cite{Savvidy:2005zm, Savvidy:2005fi}.
If successful this procedure will fix the coefficients in the linear sum.
In the next section we shall present this construction for the
rank-2 and  rank-3 tensor gauge fields from
\cite{Savvidy:2005zm,Savvidy:2005fi,Savvidy:2005ki} and
then shall turn to the consideration of the rank-4 field.

\section{Second and Third Rank Tensor Gauge Fields}
\setcounter{equation}{0}
\setcounter{footnote}{0}

In order to describe rank-two quanta,
one should introduce tensor gauge field $A_{\mu\lambda_1}(x)$
together with tensor gauge fields whose ranks would be up to
three, such as $A_{\mu}(x)$ and $A_{\mu\lambda_1\lambda_2}(x)$.
Using these fields one can construct the following Lorentz invariant quadratic
form  \cite{Savvidy:2005zm},
\begin{equation}
\CL_2
=
\tr
\bigg(
a_1 G_{\mu\nu, \rho}G_{\mu\nu, \rho}
+a_2 G_{\mu\nu}G_{\mu\nu, \rho\rho}
\bigg),
\end{equation}
where $a_1$ and $a_2$ are numerical coefficients which
should be determined by the gauge invariance of the Lagrangian.
Calculating the variation of the Lagrangian
under the gauge transformations
(\ref{gauge_trans_field_strength_0}) and
(\ref{gauge_trans_field_strength_1}),
one can get
\begin{eqnarray}
\delta\CL_2
&=&
-ig\tr
\bigg\{
2a_1
[G_{\mu\nu}, \ \xi_\rho]
G_{\mu\nu, \rho}
+2a_2
[G_{\mu\nu, \rho}, \ \xi_\rho]
G_{\mu\nu}
\bigg\}
\nonumber
\\
&=&
-ig\tr
\bigg\{
2(a_1-a_2)[G_{\mu\nu}, \ \xi_\rho]
G_{\mu\nu, \rho}
\bigg\}.
\nonumber
\end{eqnarray}
One can determine the numerical coefficients
by requiring the invariance of the Lagrangian,
$$
a_2=a_1,
$$
and find out the following Lorentz and gauge invariant Lagrangian
\cite{Savvidy:2005zm},
\begin{equation}
\CL_2
=
a_1\tr
\bigg(
G_{\mu\nu, \rho}G_{\mu\nu, \rho}
+G_{\mu\nu}G_{\mu\nu, \rho\rho}
\bigg),
\end{equation}
where the numerical coefficient $a_1$ remains arbitrary.

The second invariant Lagrangian $\CL_2'$
consists of the following three quadratic forms
\cite{Savvidy:2005fi, Savvidy:2005ki},
\begin{equation}
\CL'_2
=
\tr
\bigg(
b_1G_{\mu\nu, \rho}G_{\mu\rho, \nu}
+b_2G_{\mu\nu, \nu}G_{\mu\rho, \rho}
+b_3G_{\mu\nu}G_{\mu\rho, \nu\rho}
\bigg).
\end{equation}
Calculating the variation of the Lagrangian
under the gauge transformations
(\ref{gauge_trans_field_strength_0}) and
(\ref{gauge_trans_field_strength_1}),
one can get the following result,
\begin{eqnarray}
\delta\CL_2'
&=&
-ig\tr
\Bigg\{
2b_1
[G_{\mu\nu}, \ \xi_\rho]
G_{\mu\rho, \nu}
+2b_2
[G_{\mu\nu}, \ \xi_\nu]
G_{\mu\rho, \rho}
+b_3
\bigg(
[G_{\mu\rho, \nu}, \ \xi_\rho]
+[G_{\mu\rho, \rho}, \ \xi_\nu]
\bigg)
G_{\mu\nu}
\Bigg\}
\nonumber
\\
&=&
-ig\tr
\Bigg\{
(2b_1-b_3)
[G_{\mu\nu}, \ \xi_\rho]
G_{\mu\rho, \nu}
+(2b_2-b_3)
[G_{\mu\nu}, \ \xi_\nu]
G_{\mu\rho, \rho}
\Bigg\}.
\nonumber
\end{eqnarray}
Then one can determine the numerical coefficients as
$$
b_2=b_1, \quad
b_3=2b_1,
$$
and find the following second invariant form
\cite{Savvidy:2005fi, Savvidy:2005ki},
\begin{equation}
\CL'_2
=
b_1\tr
\bigg(
G_{\mu\nu, \rho}G_{\mu\rho, \nu}
+G_{\mu\nu, \nu}G_{\mu\rho, \rho}
+2G_{\mu\nu}G_{\mu\rho, \nu\rho}
\bigg),
\end{equation}
where the coefficient $b_1$ remains arbitrary.

For rank-three quanta,
one should introduce tensor gauge field $A_{\mu\lambda_1\lambda_2}(x)$
together with tensor gauge fields whose ranks would be up to
five, such as $A_{\mu}(x)$, $A_{\mu\lambda_1}(x)$, $\cdots$,
and $A_{\mu\lambda_1\cdots\lambda_4}(x)$.
Using these fields one can construct the following Lorentz invariant forms
\cite{Savvidy:2005zm}:
\begin{equation}
\CL_3
=
\tr
\bigg(
a_1G_{\mu\nu, \rho\sigma}G_{\mu\nu, \rho\sigma}
+a_2G_{\mu\nu\, \rho\rho}G_{\mu\nu, \sigma\sigma}
+a_3G_{\mu\nu, \rho}G_{\mu\nu, \rho\sigma\sigma}
+a_4G_{\mu\nu}G_{\mu\nu, \rho\rho\sigma\sigma}
\bigg).
\end{equation}
Calculating the variation  under the gauge
transformations (\ref{gauge_trans_field_strength_0}) and
(\ref{gauge_trans_field_strength_1}),
one can get the following result:
\begin{eqnarray}
\delta\CL_3
&=&
-ig\tr
\Bigg\{
2a_1
\bigg(
[G_{\mu\nu}, \ \xi_{\rho\sigma}]
+2[G_{\mu\nu, \rho}, \ \xi_\sigma]
\bigg)
G_{\mu\nu, \rho\sigma}
\nonumber
\\
&&
\hspace*{15mm}
+2a_2
\bigg(
[G_{\mu\nu}, \ \xi_{\rho\rho}]
+2[G_{\mu\nu, \rho}, \ \xi_\rho]
\bigg)
G_{\mu\nu, \sigma\sigma}
\nonumber
\\
&&
\hspace*{15mm}
+a_3
\Bigg(
[G_{\mu\nu}, \ \xi_\rho]
G_{\mu\nu, \rho\sigma\sigma}
\nonumber
\\
&&
\hspace*{25mm}
+\bigg(
[G_{\mu\nu}, \ \xi_{\rho\sigma\sigma}]
+2[G_{\mu\nu, \rho\sigma}, \ \xi_\sigma]
+[G_{\mu\nu, \sigma\sigma}, \ \xi_\rho]
\bigg)
G_{\mu\nu, \rho}
\Bigg)
\nonumber
\\
&&
\hspace*{15mm}
+a_4
\bigg(
4[G_{\mu\nu, \rho}, \ \xi_{\rho\sigma\sigma}]
+2[G_{\mu\nu, \rho\rho}, \ \xi_{\sigma\sigma}]
+4[G_{\mu\nu, \rho\sigma}, \ \xi_{\rho\sigma}]
+4[G_{\mu\nu, \rho\rho\sigma}, \ \xi_\sigma]
\bigg)
G_{\mu\nu}
\Bigg\}
\nonumber
\\
&=&
-ig\tr
\Bigg\{
2(2a_1-a_3)
[G_{\mu\nu, \rho}, \ \xi_\sigma]
G_{\mu\nu, \rho\sigma}
+(4a_2-a_3)
[G_{\mu\nu, \rho}, \ \xi_\rho]
G_{\mu\nu, \sigma\sigma}
\nonumber
\\
&&
\hspace*{15mm}
+(a_3-4a_4)
[G_{\mu\nu}, \ \xi_\rho]
G_{\mu\nu, \rho\sigma\sigma}
+2(a_1-2a_4)
[G_{\mu\nu}, \ \xi_{\rho\sigma}]
G_{\mu\nu, \rho\sigma}
\nonumber
\\
&&
\hspace*{15mm}
+2(a_2-a_4)
[G_{\mu\nu}, \ \xi_{\rho\rho}]
G_{\mu\nu, \sigma\sigma}
+(a_3-4a_4)
[G_{\mu\nu}, \ \xi_{\rho\sigma\sigma}]
G_{\mu\nu, \rho}
\Bigg\}
\nonumber
\end{eqnarray}
and determine the numerical coefficients consistently as
$$
a_2=\frac{1}{2}a_1, \quad
a_3=2a_1, \quad
a_4=\frac{1}{2}a_1.
$$
Thus one can get the following gauge invariant form
with the arbitrary coefficient $a_1$ \cite{Savvidy:2005zm},
\begin{equation}
\CL_3
=
a_1\tr
\bigg(
G_{\mu\nu, \rho\sigma}G_{\mu\nu, \rho\sigma}
+\frac{1}{2}G_{\mu\nu, \rho\rho}G_{\mu\nu, \sigma\sigma}
+2G_{\mu\nu, \rho}G_{\mu\nu, \rho\sigma\sigma}
+\frac{1}{2}G_{\mu\nu}G_{\mu\nu, \rho\rho\sigma\sigma}
\bigg).
\end{equation}
There are additional seven Lorentz invariant quadratic
terms \cite{Savvidy:2005fi, Savvidy:2005ki},
\begin{eqnarray}
\CL_3'
&=&
\tr
\bigg(
b_1G_{\mu\nu, \rho\sigma}G_{\mu\rho, \nu\sigma}
+b_2G_{\mu\nu, \nu\rho}G_{\mu\rho, \sigma\sigma}
+b_3G_{\mu\nu, \nu\rho}G_{\mu\sigma, \rho\sigma}
\nonumber
\\
&&
\hspace*{7mm}
+b_4G_{\mu\nu, \rho}G_{\mu\rho, \nu\sigma\sigma}
+b_5G_{\mu\nu, \rho}G_{\mu\sigma, \nu\rho\sigma}
+b_6G_{\mu\nu, \nu}G_{\mu\rho, \rho\sigma\sigma}
\nonumber
\\
&&
\hspace*{7mm}
+b_7G_{\mu\nu}G_{\mu\rho, \nu\rho\sigma\sigma}
\bigg).
\end{eqnarray}
Calculating the variation of this form under the gauge
transformations (\ref{gauge_trans_field_strength_0})
and (\ref{gauge_trans_field_strength_1}),
one can get the following result,
\begin{eqnarray}
\delta\CL_3'
&=&
-ig\tr
\Bigg\{
2b_1
\bigg(
[G_{\mu\nu}, \ \xi_{\rho\sigma}]
+[G_{\mu\nu, \rho}, \ \xi_\sigma]
+[G_{\mu\nu, \sigma}, \ \xi_\rho]
\bigg)
G_{\mu\rho, \nu\sigma}
\nonumber
\\
&&
\hspace*{15mm}
+b_2
\Bigg(
\bigg(
[G_{\mu\nu}, \ \xi_{\nu\rho}]
+[G_{\mu\nu, \nu}, \ \xi_\rho]
+[G_{\mu\nu, \rho}, \ \xi_\nu]
\bigg)
G_{\mu\rho, \sigma\sigma}
\nonumber
\\
&&
\hspace*{25mm}
+\bigg(
[G_{\mu\rho}, \ \xi_{\sigma\sigma}]
+2[G_{\mu\rho, \sigma}, \ \xi_\sigma]
\bigg)
G_{\mu\nu, \nu\rho}
\Bigg)
\nonumber
\\
&&
\hspace*{15mm}
+2b_3
\bigg(
[G_{\mu\nu}, \ \xi_{\nu\rho}]
+[G_{\mu\nu, \nu}, \ \xi_\rho]
+[G_{\mu\nu, \rho}, \ \xi_\nu]
\bigg)
G_{\mu\sigma, \rho\sigma}
\nonumber
\\
&&
\hspace*{15mm}
+b_4
\Bigg(
[G_{\mu\nu}, \ \xi_\rho]
G_{\mu\rho, \nu\sigma\sigma}
\nonumber
\\
&&
\hspace*{25mm}
+\bigg(
[G_{\mu\rho}, \ \xi_{\nu\sigma\sigma}]
+2[G_{\mu\rho, \sigma}, \ \xi_{\nu\sigma}]
+2[G_{\mu\rho, \nu\sigma}, \ \xi_\sigma]
+[G_{\mu\rho, \sigma\sigma}, \ \xi_\nu]
\bigg)
G_{\mu\nu, \rho}
\Bigg)
\nonumber
\\
&&
\hspace*{15mm}
+b_5
\Bigg(
[G_{\mu\nu}, \ \xi_\rho]
G_{\mu\sigma, \nu\rho\sigma}
\nonumber
\\
&&
\hspace*{25mm}
+\bigg(
[G_{\mu\sigma}, \ \xi_{\nu\rho\sigma}]
+[G_{\mu\sigma, \nu}, \ \xi_{\rho\sigma}]
+[G_{\mu\sigma, \sigma}, \ \xi_{\nu\rho}]
\nonumber
\\
&&
\hspace*{32mm}
+[G_{\mu\sigma, \nu\rho}, \ \xi_\sigma]
+[G_{\mu\sigma, \nu\sigma}, \ \xi_\rho]
+[G_{\mu\sigma, \rho\sigma}, \ \xi_\nu]
\bigg)
G_{\mu\nu, \rho}
\Bigg)
\nonumber
\\
&&
\hspace*{15mm}
+b_6
\Bigg(
[G_{\mu\nu}, \ \xi_\nu]
G_{\mu\rho, \rho\sigma\sigma}
\nonumber
\\
&&
\hspace*{25mm}
+\bigg(
[G_{\mu\rho}, \ \xi_{\rho\sigma\sigma}]
+2[G_{\mu\rho, \sigma}, \ \xi_{\rho\sigma}]
+2[G_{\mu\rho, \rho\sigma}, \ \xi_\sigma]
+[G_{\mu\rho, \sigma\sigma}, \ \xi_\rho]
\bigg)
G_{\mu\nu, \nu}
\Bigg)
\nonumber
\\
&&
\hspace*{15mm}
+b_7
\bigg(
[G_{\mu\rho, \nu}, \ \xi_{\rho\sigma\sigma}]
+[G_{\mu\rho, \rho}, \ \xi_{\nu\sigma\sigma}]
+2[G_{\mu\rho, \sigma}, \ \xi_{\nu\rho\sigma}]
\nonumber
\\
&&
\hspace*{25mm}
+[G_{\mu\rho, \nu\rho}, \ \xi_{\sigma\sigma}]
+2[G_{\mu\rho, \nu\sigma}, \ \xi_{\rho\sigma}]
+2[G_{\mu\rho, \rho\sigma}, \ \xi_{\nu\sigma}]
+[G_{\mu\rho, \sigma\sigma}, \ \xi_{\nu\rho}]
\nonumber
\\
&&
\hspace*{25mm}
+2[G_{\mu\rho, \nu\rho\sigma}, \ \xi_\sigma]
+[G_{\mu\rho, \rho\sigma\sigma}, \ \xi_\nu]
+[G_{\mu\rho, \nu\sigma\sigma}, \ \xi_\rho]
\bigg)
G_{\mu\nu}
\Bigg\}
\nonumber
\\
&=&
-ig\tr
\Bigg\{
2(b_1-b_4)
[G_{\mu\nu, \rho}, \ \xi_\sigma]
G_{\mu\rho, \nu\sigma}
+(2b_1-b_5)
[G_{\mu\nu, \sigma}, \ \xi_\rho]
G_{\mu\rho, \nu\sigma}
\nonumber
\\
&&
\hspace*{15mm}
+(b_2-b_6)
[G_{\mu\nu, \nu}, \ \xi_\rho]
G_{\mu\rho, \sigma\sigma}
+(b_2-b_4)
[G_{\mu\nu, \rho}, \ \xi_\nu]
G_{\mu\rho, \sigma\sigma}
\nonumber
\\
&&
\hspace*{15mm}
+(2b_2-b_5)
[G_{\mu\rho, \sigma}, \ \xi_\sigma]
G_{\mu\nu, \nu\rho}
+2(b_3-b_6)
[G_{\mu\nu, \nu}, \ \xi_\rho]
G_{\mu\sigma, \rho\sigma}
\nonumber
\\
&&
\hspace*{15mm}
+(2b_3-b_5)
[G_{\mu\nu, \rho}, \ \xi_\nu]
G_{\mu\sigma, \rho\sigma}
+(b_4-b_7)
[G_{\mu\nu}, \ \xi_\rho]
G_{\mu\rho, \nu\sigma\sigma}
\nonumber
\\
&&
\hspace*{15mm}
+(b_5-2b_7)
[G_{\mu\nu}, \ \xi_\rho]
G_{\mu\sigma, \nu\rho\sigma}
+(b_6-b_7)
[G_{\mu\nu}, \ \xi_\nu]
G_{\mu\rho, \rho\sigma\sigma}
\nonumber
\\
&&
\hspace*{15mm}
+2(b_1-b_7)
[G_{\mu\nu}, \ \xi_{\rho\sigma}]
G_{\mu\rho, \nu\sigma}
+(b_2-b_7)
[G_{\mu\nu}, \ \xi_{\nu\rho}]
G_{\mu\rho, \sigma\sigma}
\nonumber
\\
&&
\hspace*{15mm}
+(b_2-b_7)
[G_{\mu\rho}, \ \xi_{\sigma\sigma}]
G_{\mu\nu, \nu\rho}
+2(b_3-b_7)
[G_{\mu\nu}, \ \xi_{\nu\rho}]
G_{\mu\sigma, \rho\sigma}
\nonumber
\\
&&
\hspace*{15mm}
+(2b_4-b_5)
[G_{\mu\rho, \sigma}, \ \xi_{\nu\sigma}]
G_{\mu\nu, \rho}
+(b_5-2b_6)
[G_{\mu\sigma, \sigma}, \ \xi_{\nu\rho}]
G_{\mu\nu, \rho}
\nonumber
\\
&&
\hspace*{15mm}
+(b_4-b_7)
[G_{\mu\rho}, \ \xi_{\nu\sigma\sigma}]
G_{\mu\nu, \rho}
+(b_5-2b_7)
[G_{\mu\sigma}, \ \xi_{\nu\rho\sigma}]
G_{\mu\nu, \rho}
\nonumber
\\
&&
\hspace*{15mm}
+(b_6-b_7)
[G_{\mu\rho}, \ \xi_{\rho\sigma\sigma}]
G_{\mu\nu, \nu}
\Bigg\}.
\nonumber
\end{eqnarray}
This determines the numerical coefficients as
$$
b_2=b_1, \quad
b_3=b_1, \quad
b_4=b_1, \quad
b_5=2b_1, \quad
b_6=b_1, \quad
b_7=b_1,
$$
and provides the second invariant
form with the arbitrary coefficient $b_1$
\cite{Savvidy:2005fi, Savvidy:2005ki},
\begin{eqnarray}
\CL_3'
&=&
b_1\tr
\bigg(
G_{\mu\nu, \rho\sigma}G_{\mu\rho, \nu\sigma}
+G_{\mu\nu, \nu\rho}G_{\mu\rho, \sigma\sigma}
+G_{\mu\nu, \nu\rho}G_{\mu\sigma, \rho\sigma}
\nonumber
\\
&&
\hspace*{11mm}
+G_{\mu\nu, \rho}G_{\mu\rho, \nu\sigma\sigma}
+2G_{\mu\nu, \rho}G_{\mu\sigma, \nu\rho\sigma}
+G_{\mu\nu, \nu}G_{\mu\rho, \rho\sigma\sigma}
\nonumber
\\
&&
\hspace*{11mm}
+G_{\mu\nu}G_{\mu\rho, \nu\rho\sigma\sigma}
\bigg).
\end{eqnarray}
Both Lagrangians $\CL_2 +\CL_2'$ and $\CL_3 + \CL_3'$ for rank-2 and rank-3
fields coincide with the
corresponding expressions in the general Lagrangian $\CL$ (\ref{fulllagrangian2}).
In the next section we shall present an analogous construction of the
Lorentz and gauge invariant Lagrangian for the rank-4 gauge field.

\section{Fourth Rank Tensor Gauge Fields}
\setcounter{equation}{0}
\setcounter{footnote}{0}

In order to describe rank-four quanta, we shall introduce
tensor gauge field $A_{\mu\lambda_1\lambda_2\lambda_3}(x)$
together with tensor gauge fields
whose ranks would be up to seven, such as
$A_\mu(x)$, $A_{\mu\lambda_1}(x)$, $\cdots$,
and $A_{\mu\lambda_1\cdots\lambda_6}(x)$.
Using these fields, we shall get the following form
consisting of six quadratic terms:
\begin{eqnarray}
\CL_4
&=&
\tr
\bigg(
a_1G_{\mu\nu, \rho\sigma\lambda}G_{\mu\nu, \rho\sigma\lambda}
+a_2G_{\mu\nu, \rho\rho\sigma}G_{\mu\nu, \sigma\lambda\lambda}
+a_3G_{\mu\nu, \rho\sigma}G_{\mu\nu, \rho\sigma\lambda\lambda}
\nonumber
\\
&&
\hspace*{8mm}
+a_4G_{\mu\nu, \rho\rho}G_{\mu\nu, \sigma\sigma\lambda\lambda}
+a_5G_{\mu\nu, \rho}G_{\mu\nu, \rho\sigma\sigma\lambda\lambda}
+a_6G_{\mu\nu}G_{\mu\nu, \rho\rho\sigma\sigma\lambda\lambda}
\bigg).
\end{eqnarray}
In order to fix the numerical coefficients,
we shall calculate the variation of the Lagrangian under the gauge
transformations (\ref{gauge_trans_field_strength_0}) and
(\ref{gauge_trans_field_strength_1}) and get the following result:
\begin{eqnarray}
\delta\CL_4
&=&
-ig\tr
\Bigg\{
2a_1
\bigg(
[G_{\mu\nu}, \ \xi_{\rho\sigma\lambda}]
+3[G_{\mu\nu, \rho}, \ \xi_{\sigma\lambda}]
+3[G_{\mu\nu, \rho\sigma}, \ \xi_\lambda]
\bigg)
G_{\mu\nu, \rho\sigma\lambda}
\nonumber
\\
&&
\hspace*{15mm}
+2a_2
\bigg(
[G_{\mu\nu}, \ \xi_{\rho\rho\sigma}]
+2[G_{\mu\nu, \rho}, \ \xi_{\rho\sigma}]
+[G_{\mu\nu, \sigma}, \ \xi_{\rho\rho}]
\nonumber
\\
&&
\hspace*{28mm}
+[G_{\mu\nu, \rho\rho}, \ \xi_\sigma]
+2[G_{\mu\nu, \rho\sigma}, \ \xi_\rho]
\bigg)
G_{\mu\nu, \lambda\lambda\sigma}
\nonumber
\\
&&
\hspace*{15mm}
+a_3
\Bigg(
\bigg(
[G_{\mu\nu}, \ \xi_{\rho\sigma}]
+2[G_{\mu\nu, \rho}, \ \xi_\sigma]
\bigg)
G_{\mu\nu, \rho\sigma\lambda\lambda}
\nonumber
\\
&&
\hspace*{25mm}
+\bigg(
[G_{\mu\nu}, \ \xi_{\rho\sigma\lambda\lambda}]
+2[G_{\mu\nu, \rho}, \ \xi_{\sigma\lambda\lambda}]
+2[G_{\mu\nu, \lambda}, \ \xi_{\rho\sigma\lambda}]
\nonumber
\\
&&
\hspace*{32mm}
+[G_{\mu\nu, \lambda\lambda}, \ \xi_{\rho\sigma}]
+2[G_{\mu\nu, \rho\sigma\lambda}, \ \xi_\lambda]
+2[G_{\mu\nu, \rho\lambda\lambda}, \ \xi_\sigma]
\bigg)
G_{\mu\nu, \rho\sigma}
\Bigg)
\nonumber
\\
&&
\hspace*{15mm}
+a_4
\Bigg(
\bigg(
[G_{\mu\nu}, \ \xi_{\rho\rho}]
+2[G_{\mu\nu, \rho}, \ \xi_\rho]
\bigg)
G_{\mu\nu, \sigma\sigma\lambda\lambda}
\nonumber
\\
&&
\hspace*{25mm}
+\bigg(
[G_{\mu\nu}, \ \xi_{\sigma\sigma\lambda\lambda}]
+4[G_{\mu\nu, \sigma}, \ \xi_{\sigma\lambda\lambda}]
\nonumber
\\
&&
\hspace*{32mm}
+4[G_{\mu\nu, \sigma\lambda}, \ \xi_{\sigma\lambda}]
+4[G_{\mu\nu, \sigma\sigma\lambda}, \ \xi_\lambda]
\bigg)
G_{\mu\nu, \rho\rho}
\Bigg)
\nonumber
\\
&&
\hspace*{15mm}
+a_5
\Bigg(
[G_{\mu\nu}, \ \xi_\rho]
G_{\mu\nu, \rho\sigma\sigma\lambda\lambda}
\nonumber
\\
&&
\hspace*{25mm}
+\bigg(
[G_{\mu\nu}, \ \xi_{\rho\sigma\sigma\lambda\lambda}]
+4[G_{\mu\nu, \rho\sigma}, \ \xi_{\sigma\lambda\lambda}]
+2[G_{\mu\nu, \sigma\sigma}, \ \xi_{\rho\lambda\lambda}]
\nonumber
\\
&&
\hspace*{32mm}
+4[G_{\mu\nu, \sigma\lambda}, \ \xi_{\rho\sigma\lambda}]
+2[G_{\mu\nu, \rho\sigma\sigma}, \ \xi_{\lambda\lambda}]
+4[G_{\mu\nu, \rho\sigma\lambda}, \ \xi_{\sigma\lambda}]
\nonumber
\\
&&
\hspace*{32mm}
+4[G_{\mu\nu, \sigma\sigma\lambda}, \ \xi_{\rho\lambda}]
+4[G_{\mu\nu, \rho\sigma\sigma\lambda}, \ \xi_\lambda]
+[G_{\mu\nu, \sigma\sigma\lambda\lambda}, \ \xi_\rho]
\bigg)
G_{\mu\nu, \rho}
\Bigg)
\nonumber
\\
&&
\hspace*{15mm}
+a_6
\bigg(
6[G_{\mu\nu, \rho}, \ \xi_{\rho\sigma\sigma\lambda\lambda}]
+3[G_{\mu\nu, \rho\rho}, \ \xi_{\sigma\sigma\lambda\lambda}]
+12[G_{\mu\nu, \rho\sigma}, \ \xi_{\rho\sigma\lambda\lambda}]
\nonumber
\\
&&
\hspace*{25mm}
+12[G_{\mu\nu, \rho\rho\sigma}, \ \xi_{\sigma\lambda\lambda}]
+8[G_{\mu\nu, \rho\sigma\lambda}, \ \xi_{\rho\sigma\lambda}]
+3[G_{\mu\nu, \rho\rho\sigma\sigma}, \ \xi_{\lambda\lambda}]
\nonumber
\\
&&
\hspace*{25mm}
+12[G_{\mu\nu, \rho\rho\sigma\lambda}, \ \xi_{\sigma\lambda}]
+6[G_{\mu\nu, \rho\rho\sigma\sigma\lambda}, \ \xi_\lambda]
\bigg)
G_{\mu\nu}
\Bigg\}
\nonumber
\\
&=&
-ig\tr
\Bigg\{
2(3a_1-a_3)
[G_{\mu\nu, \rho\sigma}, \ \xi_\lambda]
G_{\mu\nu, \rho\sigma\lambda}
+2(a_2-2a_4)
[G_{\mu\nu, \rho\rho}, \ \xi_\sigma]
G_{\mu\nu, \lambda\lambda\sigma}
\nonumber
\\
&&
\hspace*{15mm}
+2(2a_2-a_3)
[G_{\mu\nu, \rho\sigma}, \ \xi_\rho]
G_{\mu\nu, \lambda\lambda\sigma}
+2(a_3-2a_4)
[G_{\mu\nu, \rho}, \ \xi_\sigma]
G_{\mu\nu, \rho\sigma\lambda\lambda}
\nonumber
\\
&&
\hspace*{15mm}
+(2a_4-a_5)
[G_{\mu\nu, \rho}, \ \xi_\rho]
G_{\mu\nu, \sigma\sigma\lambda\lambda}
+(a_5-6a_6)
[G_{\mu\nu}, \ \xi_\rho]
G_{\mu\nu, \rho\sigma\sigma\lambda\lambda}
\nonumber
\\
&&
\hspace*{15mm}
+2(3a_1-2a_5)
[G_{\mu\nu, \rho}, \ \xi_{\sigma\lambda}]
G_{\mu\nu, \rho\sigma\lambda}
+4(a_2-a_5)
[G_{\mu\nu, \rho}, \ \xi_{\rho\sigma}]
G_{\mu\nu, \lambda\lambda\sigma}
\nonumber
\\
&&
\hspace*{15mm}
+2(a_2-a_5)
[G_{\mu\nu, \sigma}, \ \xi_{\rho\rho}]
G_{\mu\nu, \lambda\lambda\sigma}
+(a_3-12a_6)
[G_{\mu\nu}, \ \xi_{\rho\sigma}]
G_{\mu\nu, \rho\sigma\lambda\lambda}
\nonumber
\\
&&
\hspace*{15mm}
+(a_3-4a_4)
[G_{\mu\nu, \lambda\lambda}, \ \xi_{\rho\sigma}]
G_{\mu\nu, \rho\sigma}
+(a_4-3a_6)
[G_{\mu\nu}, \ \xi_{\rho\rho}]
G_{\mu\nu, \sigma\sigma\lambda\lambda}
\nonumber
\\
&&
\hspace*{15mm}
+2(a_1-4a_6)
[G_{\mu\nu}, \ \xi_{\rho\sigma\lambda}]
G_{\mu\nu, \rho\sigma\lambda}
+2(a_2-6a_6)
[G_{\mu\nu}, \ \xi_{\rho\rho\sigma}]
G_{\mu\nu, \lambda\lambda\sigma}
\nonumber
\\
&&
\hspace*{15mm}
+2(a_3-2a_5)
[G_{\mu\nu, \sigma}, \ \xi_{\rho\lambda\lambda}]
G_{\mu\nu, \rho\sigma}
+2(a_3-2a_5)
[G_{\mu\nu, \lambda}, \ \xi_{\rho\sigma\lambda}]
G_{\mu\nu, \rho\sigma}
\nonumber
\\
&&
\hspace*{15mm}
+2(2a_4-a_5)
[G_{\mu\nu, \sigma}, \ \xi_{\sigma\lambda\lambda}]
G_{\mu\nu, \rho\rho}
+(a_3-12a_6)
[G_{\mu\nu}, \ \xi_{\rho\sigma\lambda\lambda}]
G_{\mu\nu, \rho\sigma}
\nonumber
\\
&&
\hspace*{15mm}
+(a_4-3a_6)
[G_{\mu\nu}, \ \xi_{\sigma\sigma\lambda\lambda}]
G_{\mu\nu, \rho\rho}
+(a_5-6a_6)
[G_{\mu\nu}, \ \xi_{\rho\sigma\sigma\lambda\lambda}]
G_{\mu\nu, \rho}
\Bigg\}.
\nonumber
\end{eqnarray}
Then we can determine the coefficients consistently as
$$
a_2=\frac{3}{2}a_1, \quad
a_3=3a_1, \quad
a_4=\frac{3}{4}a_1, \quad
a_5=\frac{3}{2}a_1, \quad
a_6=\frac{1}{4}a_1,
$$
and come to the following gauge invariant Lagrangian \cite{Savvidy:2005zm}:
\begin{eqnarray}
\CL_4
&=&
a_1\tr
\bigg(
G_{\mu\nu, \rho\sigma\lambda}G_{\mu\nu, \rho\sigma\lambda}
+\frac{3}{2}
G_{\mu\nu, \rho\rho\sigma}G_{\mu\nu, \sigma\lambda\lambda}
+3
G_{\mu\nu, \rho\sigma}G_{\mu\nu, \rho\sigma\lambda\lambda}
\nonumber
\\
&&
\hspace*{12mm}
+\frac{3}{4}
G_{\mu\nu, \rho\rho}G_{\mu\nu, \sigma\sigma\lambda\lambda}
+\frac{3}{2}
G_{\mu\nu, \rho}G_{\mu\nu, \rho\sigma\sigma\lambda\lambda}
+\frac{1}{4}
G_{\mu\nu}G_{\mu\nu, \rho\rho\sigma\sigma\lambda\lambda}
\bigg).
\end{eqnarray}

Now let us proceed in order to define the second invariant Lagrangian
$\CL_4'$.
There are fourteen Lorentz invariant quadratic terms,
\begin{eqnarray}
\CL_4'
&=&
\tr
\bigg(
b_1
G_{\mu\nu, \rho\sigma\lambda}G_{\mu\rho, \nu\sigma\lambda}
+b_2
G_{\mu\nu, \rho\sigma\sigma}G_{\mu\rho, \nu\lambda\lambda}
+b_3
G_{\mu\nu, \rho\sigma\sigma}G_{\mu\lambda, \nu\rho\lambda}
+b_4
G_{\mu\nu, \nu\rho\sigma}G_{\mu\lambda, \rho\sigma\lambda}
\nonumber
\\
&&
\hspace*{7mm}
+b_5
G_{\mu\nu, \nu\rho\rho}G_{\mu\sigma, \sigma\lambda\lambda}
+b_6
G_{\mu\nu, \rho\sigma}G_{\mu\rho, \nu\sigma\lambda\lambda}
+b_7
G_{\mu\nu, \rho\sigma}G_{\mu\lambda, \nu\rho\sigma\lambda}
+b_8
G_{\mu\nu, \rho\rho}G_{\mu\sigma, \nu\sigma\lambda\lambda}
\nonumber
\\
&&
\hspace*{7mm}
+b_9
G_{\mu\nu, \nu\rho}G_{\mu\rho, \lambda\lambda\sigma\sigma}
+b_{10}
G_{\mu\nu, \nu\rho}G_{\mu\lambda, \rho\lambda\sigma\sigma}
+b_{11}
G_{\mu\nu, \rho}G_{\mu\rho, \nu\lambda\lambda\sigma\sigma}
+b_{12}
G_{\mu\nu, \rho}G_{\mu\lambda, \nu\rho\lambda\sigma\sigma}
\nonumber
\\
&&
\hspace*{7mm}
+b_{13}
G_{\mu\nu, \nu}G_{\mu\rho, \rho\sigma\sigma\lambda\lambda}
+b_{14}
G_{\mu\nu}G_{\mu\rho, \nu\rho\sigma\sigma\lambda\lambda}
\bigg).
\end{eqnarray}
After calculation for the variation of the Lagrangian
under the gauge transformations (\ref{gauge_trans_field_strength_0}) and
(\ref{gauge_trans_field_strength_1}),
we arrive at the following result,
\begin{eqnarray}
\delta\CL_4'
&=&
-ig\tr
\Bigg\{
2b_1
\bigg(
[G_{\mu\nu}, \ \xi_{\rho\sigma\lambda}]
+[G_{\mu\nu, \rho}, \ \xi_{\sigma\lambda}]
+2[G_{\mu\nu, \sigma}, \ \xi_{\rho\lambda}]
\nonumber
\\
&&
\hspace*{24mm}
+2[G_{\mu\nu, \rho\sigma}, \ \xi_\lambda]
+[G_{\mu\nu, \sigma\lambda}, \ \xi_\rho]
\bigg)
G_{\mu\rho, \nu\sigma\lambda}
\nonumber
\\
&&
\hspace*{15mm}
+2b_2
\bigg(
[G_{\mu\nu}, \ \xi_{\rho\sigma\sigma}]
+[G_{\mu\nu, \rho}, \ \xi_{\sigma\sigma}]
+2[G_{\mu\nu, \sigma}, \ \xi_{\rho\sigma}]
\nonumber
\\
&&
\hspace*{28mm}
+2[G_{\mu\nu, \rho\sigma}, \ \xi_\sigma]
+[G_{\mu\nu, \sigma\sigma}, \ \xi_\rho]
\bigg)
G_{\mu\rho, \nu\lambda\lambda}
\nonumber
\\
&&
\hspace*{15mm}
+b_3
\Bigg(
\bigg(
[G_{\mu\nu}, \ \xi_{\rho\sigma\sigma}]
+[G_{\mu\nu, \rho}, \ \xi_{\sigma\sigma}]
+2[G_{\mu\nu, \sigma}, \ \xi_{\rho\sigma}]
\nonumber
\\
&&
\hspace*{28mm}
+2[G_{\mu\nu, \rho\sigma}, \ \xi_\sigma]
+[G_{\mu\nu, \sigma\sigma}, \ \xi_\rho]
\bigg)
G_{\mu\lambda, \nu\rho\lambda}
\nonumber
\\
&&
\hspace*{25mm}
+
\bigg(
[G_{\mu\lambda}, \ \xi_{\nu\rho\lambda}]
+[G_{\mu\lambda, \nu}, \ \xi_{\rho\lambda}]
+[G_{\mu\lambda, \rho}, \ \xi_{\nu\lambda}]
+[G_{\mu\lambda, \lambda}, \ \xi_{\nu\rho}]
\nonumber
\\
&&
\hspace*{31mm}
+[G_{\mu\lambda, \nu\rho}, \ \xi_\lambda]
+[G_{\mu\lambda, \nu\lambda}, \ \xi_\rho]
+[G_{\mu\lambda, \rho\lambda}, \ \xi_\nu]
\bigg)
G_{\mu\nu, \rho\sigma\sigma}
\Bigg)
\nonumber
\\
&&
\hspace*{15mm}
+2b_4
\bigg(
[G_{\mu\nu}, \ \xi_{\nu\rho\sigma}]
+[G_{\mu\nu, \nu}, \ \xi_{\rho\sigma}]
+2[G_{\mu\nu, \rho}, \ \xi_{\nu\sigma}]
\nonumber
\\
&&
\hspace*{28mm}
+2[G_{\mu\nu, \nu\rho}, \ \xi_\sigma]
+[G_{\mu\nu, \rho\sigma}, \ \xi_\nu]
\bigg)
G_{\mu\lambda, \rho\sigma\lambda}
\nonumber
\\
&&
\hspace*{15mm}
+2b_5
\bigg(
[G_{\mu\nu}, \ \xi_{\nu\rho\rho}]
+[G_{\mu\nu, \nu}, \ \xi_{\rho\rho}]
+2[G_{\mu\nu, \rho}, \ \xi_{\nu\rho}]
\nonumber
\\
&&
\hspace*{28mm}
+2[G_{\mu\nu, \nu\rho}, \ \xi_\rho]
+[G_{\mu\nu, \rho\rho}, \ \xi_\nu]
\bigg)
G_{\mu\sigma, \sigma\lambda\lambda}
\nonumber
\\
&&
\hspace*{15mm}
+b_6
\Bigg(
\bigg(
[G_{\mu\nu}, \ \xi_{\rho\sigma}]
+[G_{\mu\nu, \rho}, \ \xi_\sigma]
+[G_{\mu\nu, \sigma}, \ \xi_\rho]
\bigg)
G_{\mu\rho, \nu\sigma\lambda\lambda}
\nonumber
\\
&&
\hspace*{25mm}
+
\bigg(
[G_{\mu\rho}, \ \xi_{\nu\sigma\lambda\lambda}]
+[G_{\mu\rho, \nu}, \ \xi_{\sigma\lambda\lambda}]
+[G_{\mu\rho, \sigma}, \ \xi_{\nu\lambda\lambda}]
\nonumber
\\
&&
\hspace*{31mm}
+2[G_{\mu\rho, \lambda}, \ \xi_{\nu\sigma\lambda}]
+2[G_{\mu\rho, \sigma\lambda}, \ \xi_{\nu\lambda}]
+[G_{\mu\rho, \lambda\lambda}, \ \xi_{\nu\sigma}]
\nonumber
\\
&&
\hspace*{31mm}
+2[G_{\mu\rho, \nu\sigma\lambda}, \ \xi_\lambda]
+[G_{\mu\rho, \nu\lambda\lambda}, \ \xi_\sigma]
+[G_{\mu\rho, \sigma\lambda\lambda}, \ \xi_\nu]
\bigg)
G_{\mu\nu, \rho\sigma}
\Bigg)
\nonumber
\\
&&
\hspace*{15mm}
+b_7
\Bigg(
\bigg(
[G_{\mu\nu}, \ \xi_{\rho\sigma}]
+2[G_{\mu\nu, \rho}, \ \xi_\sigma]
\bigg)
G_{\mu\lambda, \nu\rho\sigma\lambda}
\nonumber
\\
&&
\hspace*{25mm}
+
\bigg(
[G_{\mu\lambda}, \ \xi_{\nu\rho\sigma\lambda}]
+[G_{\mu\lambda, \nu}, \ \xi_{\rho\sigma\lambda}]
+2[G_{\mu\lambda, \rho}, \ \xi_{\nu\lambda\sigma}]
+[G_{\mu\lambda, \lambda}, \ \xi_{\nu\rho\sigma}]
\nonumber
\\
&&
\hspace*{31mm}
+2[G_{\mu\lambda, \nu\rho}, \ \xi_{\sigma\lambda}]
+[G_{\mu\lambda, \nu\lambda}, \ \xi_{\rho\sigma}]
+2[G_{\mu\lambda, \rho\lambda}, \ \xi_{\nu\sigma}]
\nonumber
\\
&&
\hspace*{31mm}
+[G_{\mu\lambda, \nu\rho\sigma}, \ \xi_\lambda]
+2[G_{\mu\lambda, \nu\rho\lambda}, \ \xi_\sigma]
+[G_{\mu\lambda, \rho\sigma\lambda}, \ \xi_\nu]
\bigg)
G_{\mu\nu, \rho\sigma}
\Bigg)
\nonumber
\\
&&
\hspace*{15mm}
+b_8
\Bigg(
\bigg(
[G_{\mu\nu}, \ \xi_{\rho\rho}]
+2[G_{\mu\nu, \rho}, \ \xi_\rho]
\bigg)
G_{\mu\sigma, \nu\sigma\lambda\lambda}
\nonumber
\\
&&
\hspace*{25mm}
+
\bigg(
[G_{\mu\sigma}, \ \xi_{\nu\sigma\lambda\lambda}]
+[G_{\mu\sigma, \nu}, \ \xi_{\sigma\lambda\lambda}]
+[G_{\mu\sigma, \sigma}, \ \xi_{\nu\lambda\lambda}]
+2[G_{\mu\sigma, \lambda}, \ \xi_{\nu\sigma\lambda}]
\nonumber
\\
&&
\hspace*{31mm}
+[G_{\mu\sigma, \nu\sigma}, \ \ \xi_{\lambda\lambda}]
+2[G_{\mu\sigma, \nu\lambda}, \ \xi_{\sigma\lambda}]
+2[G_{\mu\sigma, \sigma\lambda}, \ \xi_{\nu\lambda}]
\nonumber
\\
&&
\hspace*{31mm}
+2[G_{\mu\sigma, \nu\sigma\lambda}, \ \xi_\lambda]
+[G_{\mu\sigma, \nu\lambda\lambda}, \ \xi_\sigma]
+[G_{\mu\sigma, \sigma\lambda\lambda}, \ \xi_\nu]
\bigg)
G_{\mu\nu, \rho\rho}
\Bigg)
\nonumber
\\
&&
\hspace*{15mm}
+b_9
\Bigg(
\bigg(
[G_{\mu\nu}, \ \xi_{\nu\rho}]
+[G_{\mu\nu, \nu}, \ \xi_\rho]
+[G_{\mu\nu, \rho}, \ \xi_\nu]
\bigg)
G_{\mu\rho, \lambda\lambda\sigma\sigma}
\nonumber
\\
&&
\hspace*{25mm}
+
\bigg(
[G_{\mu\rho}, \ \xi_{\lambda\lambda\sigma\sigma}]
+4[G_{\mu\rho, \lambda}, \ \xi_{\lambda\sigma\sigma}]
+2[G_{\mu\rho, \lambda\lambda}, \ \xi_{\sigma\sigma}]
\nonumber
\\
&&
\hspace*{31mm}
+4[G_{\mu\rho, \lambda\sigma}, \ \xi_{\lambda\sigma}]
+4[G_{\mu\rho, \lambda\lambda\sigma}, \ \xi_\sigma]
\bigg)
G_{\mu\nu, \nu\rho}
\Bigg)
\nonumber
\\
&&
\hspace*{15mm}
+b_{10}
\Bigg(
\bigg(
[G_{\mu\nu}, \ \xi_{\nu\rho}]
+[G_{\mu\nu, \nu}, \ \xi_\rho]
+[G_{\mu\nu, \rho}, \ \xi_\nu]
\bigg)
G_{\mu\lambda, \rho\lambda\sigma\sigma}
\nonumber
\\
&&
\hspace*{25mm}
+
\bigg(
[G_{\mu\lambda}, \ \xi_{\rho\lambda\sigma\sigma}]
+[G_{\mu\lambda, \rho}, \ \xi_{\lambda\sigma\sigma}]
+[G_{\mu\lambda, \lambda}, \ \xi_{\rho\sigma\sigma}]
\nonumber
\\
&&
\hspace*{31mm}
+2[G_{\mu\lambda, \sigma}, \ \xi_{\rho\lambda\sigma}]
+2[G_{\mu\lambda, \rho\sigma}, \ \xi_{\lambda\sigma}]
+[G_{\mu\lambda, \sigma\sigma}, \ \xi_{\rho\lambda}]
\nonumber
\\
&&
\hspace*{31mm}
+2[G_{\mu\lambda, \rho\lambda\sigma}, \ \xi_\sigma]
+[G_{\mu\lambda, \rho\sigma\sigma}, \ \xi_\lambda]
+[G_{\mu\lambda, \lambda\sigma\sigma}, \ \xi_\rho]
\bigg)
G_{\mu\nu, \nu\rho}
\Bigg)
\nonumber
\\
&&
\hspace*{15mm}
+b_{11}
\Bigg(
[G_{\mu\nu}, \ \xi_\rho]
G_{\mu\rho, \nu\lambda\lambda\sigma\sigma}
\nonumber
\\
&&
\hspace*{25mm}
+
\bigg(
[G_{\mu\rho}, \ \xi_{\nu\lambda\lambda\sigma\sigma}]
+4[G_{\mu\rho, \lambda}, \ \xi_{\nu\lambda\sigma\sigma}]
+4[G_{\mu\rho, \nu\lambda}, \ \xi_{\lambda\sigma\sigma}]
+2[G_{\mu\rho, \lambda\lambda}, \ \xi_{\nu\sigma\sigma}]
\nonumber
\\
&&
\hspace*{31mm}
+4[G_{\mu\rho, \lambda\sigma}, \ \xi_{\nu\lambda\sigma}]
+2[G_{\mu\rho, \nu\lambda\lambda}, \ \xi_{\sigma\sigma}]
+4[G_{\mu\rho, \nu\lambda\sigma}, \ \xi_{\lambda\sigma}]
\nonumber
\\
&&
\hspace*{31mm}
+4[G_{\mu\rho, \lambda\lambda\sigma}, \ \xi_{\nu\sigma}]
+4[G_{\mu\rho, \nu\lambda\lambda\sigma}, \ \xi_\sigma]
+[G_{\mu\rho, \lambda\lambda\sigma\sigma}, \ \xi_\nu]
\bigg)
G_{\mu\nu, \rho}
\Bigg)
\nonumber
\\
&&
\hspace*{15mm}
+b_{12}
\Bigg(
[G_{\mu\nu}, \ \xi_\rho]
G_{\mu\lambda, \nu\rho\lambda\sigma\sigma}
\nonumber
\\
&&
\hspace*{25mm}
+\bigg(
[G_{\mu\lambda}, \ \xi_{\nu\rho\lambda\sigma\sigma}]
+[G_{\mu\lambda, \nu}, \ \xi_{\rho\lambda\sigma\sigma}]
+[G_{\mu\lambda, \lambda}, \ \xi_{\nu\rho\sigma\sigma}]
+[G_{\mu\lambda, \nu\rho}, \ \xi_{\lambda\sigma\sigma}]
\nonumber
\\
&&
\hspace*{31mm}
+[G_{\mu\lambda, \nu\lambda}, \ \xi_{\rho\sigma\sigma}]
+2[G_{\mu\lambda, \nu\sigma}, \ \xi_{\rho\lambda\sigma}]
+[G_{\mu\lambda, \rho\lambda}, \ \xi_{\nu\sigma\sigma}]
+2[G_{\mu\lambda, \rho\sigma}, \ \xi_{\nu\lambda\sigma}]
\nonumber
\\
&&
\hspace*{31mm}
+2[G_{\mu\lambda, \lambda\sigma}, \ \xi_{\nu\rho\sigma}]
+[G_{\mu\lambda, \sigma\sigma}, \ \xi_{\nu\rho\lambda}]
+[G_{\mu\lambda, \nu\rho\lambda}, \ \xi_{\sigma\sigma}]
+2[G_{\mu\lambda, \nu\rho\sigma}, \ \xi_{\lambda\sigma}]
\nonumber
\\
&&
\hspace*{31mm}
+2[G_{\mu\lambda, \nu\lambda\sigma}, \ \xi_{\rho\sigma}]
+[G_{\mu\lambda, \nu\sigma\sigma}, \ \xi_{\rho\lambda}]
+2[G_{\mu\lambda, \rho\lambda\sigma}, \ \xi_{\nu\sigma}]
+[G_{\mu\lambda, \rho\sigma\sigma}, \ \xi_{\nu\lambda}]
\nonumber
\\
&&
\hspace*{31mm}
+[G_{\mu\lambda, \lambda\sigma\sigma}, \ \xi_{\nu\rho}]
+2[G_{\mu\lambda, \nu\rho\lambda\sigma}, \ \xi_\sigma]
+[G_{\mu\lambda, \nu\rho\sigma\sigma}, \ \xi_\lambda]
+[G_{\mu\lambda, \nu\lambda\sigma\sigma}, \ \xi_\rho]
\nonumber
\\
&&
\hspace*{31mm}
+[G_{\mu\lambda, \lambda\sigma\sigma\rho}, \ \xi_\nu]
\bigg)
G_{\mu\nu, \rho}
\Bigg)
\nonumber
\\
&&
\hspace*{15mm}
+b_{13}
\Bigg(
[G_{\mu\nu}, \ \xi_\nu]
G_{\mu\rho, \rho\sigma\sigma\lambda\lambda}
\nonumber
\\
&&
\hspace*{25mm}
+
\bigg(
[G_{\mu\rho}, \ \xi_{\rho\sigma\sigma\lambda\lambda}]
+4[G_{\mu\rho, \sigma}, \ \xi_{\rho\sigma\lambda\lambda}]
+4[G_{\mu\rho, \rho\sigma}, \ \xi_{\sigma\lambda\lambda}]
+2[G_{\mu\rho, \sigma\sigma}, \ \xi_{\rho\lambda\lambda}]
\nonumber
\\
&&
\hspace*{31mm}
+4[G_{\mu\rho, \sigma\lambda}, \ \xi_{\rho\sigma\lambda}]
+2[G_{\mu\rho, \rho\sigma\sigma}, \ \xi_{\lambda\lambda}]
+4[G_{\mu\rho, \rho\sigma\lambda}, \ \xi_{\sigma\lambda}]
\nonumber
\\
&&
\hspace*{31mm}
+4[G_{\mu\rho, \sigma\sigma\lambda}, \ \xi_{\rho\lambda}]
+4[G_{\mu\rho, \rho\sigma\sigma\lambda}, \ \xi_\lambda]
+[G_{\mu\rho, \sigma\sigma\lambda\lambda}, \ \xi_\rho]
\bigg)
G_{\mu\nu, \nu}
\Bigg)
\nonumber
\\
&&
\hspace*{15mm}
+b_{14}
\bigg(
[G_{\mu\rho, \nu}, \ \xi_{\rho\sigma\sigma\lambda\lambda}]
+[G_{\mu\rho, \rho}, \ \xi_{\nu\sigma\sigma\lambda\lambda}]
+4[G_{\mu\rho, \sigma}, \ \xi_{\nu\rho\sigma\lambda\lambda}]
+[G_{\mu\rho, \nu\rho}, \ \xi_{\sigma\sigma\lambda\lambda}]
\nonumber
\\
&&
\hspace*{27mm}
+4[G_{\mu\rho, \nu\sigma}, \ \xi_{\rho\sigma\lambda\lambda}]
+4[G_{\mu\rho, \rho\sigma}, \ \xi_{\nu\sigma\lambda\lambda}]
+2[G_{\mu\rho, \sigma\sigma}, \ \xi_{\nu\rho\lambda\lambda}]
\nonumber
\\
&&
\hspace*{27mm}
+4[G_{\mu\rho, \sigma\lambda}, \ \xi_{\nu\rho\sigma\lambda}]
+4[G_{\mu\rho, \nu\rho\sigma}, \ \xi_{\sigma\lambda\lambda}]
+2[G_{\mu\rho, \nu\sigma\sigma}, \ \xi_{\rho\lambda\lambda}]
\nonumber
\\
&&
\hspace*{27mm}
+4[G_{\mu\rho, \nu\sigma\lambda}, \ \xi_{\rho\sigma\lambda}]
+2[G_{\mu\rho, \rho\sigma\sigma}, \ \xi_{\nu\lambda\lambda}]
+4[G_{\mu\rho, \rho\sigma\lambda}, \ \xi_{\nu\sigma\lambda}]
\nonumber
\\
&&
\hspace*{27mm}
+4[G_{\mu\rho, \sigma\sigma\lambda}, \ \xi_{\nu\rho\lambda}]
+2[G_{\mu\rho, \nu\rho\sigma\sigma}, \ \xi_{\lambda\lambda}]
+4[G_{\mu\rho, \nu\rho\sigma\lambda}, \ \xi_{\sigma\lambda}]
\nonumber
\\
&&
\hspace*{27mm}
+4[G_{\mu\rho, \nu\sigma\sigma\lambda}, \ \xi_{\rho\lambda}]
+4[G_{\mu\rho, \rho\sigma\sigma\lambda}, \ \xi_{\nu\lambda}]
+[G_{\mu\rho, \sigma\sigma\lambda\lambda}, \ \xi_{\nu\rho}]
\nonumber
\\
&&
\hspace*{27mm}
+4[G_{\mu\rho, \nu\rho\sigma\sigma\lambda}, \ \xi_\lambda]
+[G_{\mu\rho, \nu\sigma\sigma\lambda\lambda}, \ \xi_\rho]
+[G_{\mu\rho, \rho\sigma\sigma\lambda\lambda}, \ \xi_\nu]
\bigg)
G_{\mu\nu}
\Bigg\}
\nonumber
\\
&=&
-ig\tr
\Bigg\{
2(2b_1-b_6)
[G_{\mu\nu, \rho\sigma}, \ \xi_\lambda]
G_{\mu\rho, \nu\sigma\lambda}
+(2b_1-b_7)
[G_{\mu\nu, \sigma\lambda}, \ \xi_\rho]
G_{\mu\rho, \nu\sigma\lambda}
\nonumber
\\
&&
\hspace*{15mm}
+(4b_2-b_6)
[G_{\mu\nu, \rho\sigma}, \ \xi_\sigma]
G_{\mu\rho, \nu\lambda\lambda}
+(2b_2-b_8)
[G_{\mu\nu, \sigma\sigma}, \ \xi_\rho]
G_{\mu\rho, \nu\lambda\lambda}
\nonumber
\\
&&
\hspace*{15mm}
+2(b_3-b_7)
[G_{\mu\nu, \rho\sigma}, \ \xi_\sigma]
G_{\mu\lambda, \nu\rho\lambda}
+(b_3-2b_8)
[G_{\mu\nu, \sigma\sigma}, \ \xi_\rho]
G_{\mu\lambda, \nu\rho\lambda}
\nonumber
\\
&&
\hspace*{15mm}
+(b_3-b_6)
[G_{\mu\lambda, \nu\rho}, \ \xi_\lambda]
G_{\mu\nu, \rho\sigma\sigma}
+(b_3-4b_9)
[G_{\mu\lambda, \nu\lambda}, \ \xi_\rho]
G_{\mu\nu, \rho\sigma\sigma}
\nonumber
\\
&&
\hspace*{15mm}
+(b_3-b_{10})
[G_{\mu\lambda, \rho\lambda}, \ \xi_\nu]
G_{\mu\nu, \rho\sigma\sigma}
+2(2b_4-b_{10})
[G_{\mu\nu, \nu\rho}, \ \xi_\sigma]
G_{\mu\lambda, \rho\sigma\lambda}
\nonumber
\\
&&
\hspace*{15mm}
+(2b_4-b_7)
[G_{\mu\nu, \rho\sigma}, \ \xi_\nu]
G_{\mu\lambda, \rho\sigma\lambda}
+(4b_5-b_{10})
[G_{\mu\nu, \nu\rho}, \ \xi_\rho]
G_{\mu\sigma, \sigma\lambda\lambda}
\nonumber
\\
&&
\hspace*{15mm}
+(2b_5-b_8)
[G_{\mu\nu, \rho\rho}, \ \xi_\nu]
G_{\mu\sigma, \sigma\lambda\lambda}
+(b_6-4b_{11})
[G_{\mu\nu, \rho}, \ \xi_\sigma]
G_{\mu\rho, \nu\sigma\lambda\lambda}
\nonumber
\\
&&
\hspace*{15mm}
+(b_6-b_{12})
[G_{\mu\nu, \sigma}, \ \xi_\rho]
G_{\mu\rho, \nu\sigma\lambda\lambda}
+2(b_7-b_{12})
[G_{\mu\nu, \rho}, \ \xi_\sigma]
G_{\mu\lambda, \nu\rho\sigma\lambda}
\nonumber
\\
&&
\hspace*{15mm}
+(2b_8-b_{12})
[G_{\mu\nu, \rho}, \ \xi_\rho]
G_{\mu\sigma, \nu\sigma\lambda\lambda}
+(b_9-b_{13})
[G_{\mu\nu, \nu}, \ \xi_\rho]
G_{\mu\rho, \lambda\lambda\sigma\sigma}
\nonumber
\\
&&
\hspace*{15mm}
+(b_9-b_{11})
[G_{\mu\nu, \rho}, \ \xi_\nu]
G_{\mu\rho, \lambda\lambda\sigma\sigma}
+(b_{10}-4b_{13})
[G_{\mu\nu, \nu}, \ \xi_\rho]
G_{\mu\lambda, \rho\lambda\sigma\sigma}
\nonumber
\\
&&
\hspace*{15mm}
+(b_{10}-b_{12})
[G_{\mu\nu, \rho}, \ \xi_\nu]
G_{\mu\lambda, \rho\lambda\sigma\sigma}
+(b_{11}-b_{14})
[G_{\mu\nu}, \ \xi_\rho]
G_{\mu\rho, \nu\lambda\lambda\sigma\sigma}
\nonumber
\\
&&
\hspace*{15mm}
+(b_{12}-4b_{14})
[G_{\mu\nu}, \ \xi_\rho]
G_{\mu\lambda, \nu\rho\lambda\sigma\sigma}
+(b_{13}-b_{14})
[G_{\mu\nu}, \ \xi_\nu]
G_{\mu\rho, \rho\sigma\sigma\lambda\lambda}
\nonumber
\\
&&
\hspace*{15mm}
+2(b_1-2b_{11})
[G_{\mu\nu, \rho}, \ \xi_{\sigma\lambda}]
G_{\mu\rho, \nu\sigma\lambda}
+2(2b_1-b_{12})
[G_{\mu\nu, \sigma}, \ \xi_{\rho\lambda}]
G_{\mu\rho. \nu\sigma\lambda}
\nonumber
\\
&&
\hspace*{15mm}
+2(b_2-b_{11})
[G_{\mu\nu, \rho}, \ \xi_{\sigma\sigma}]
G_{\mu\rho, \nu\lambda\lambda}
+(4b_2-b_{12})
[G_{\mu\nu, \sigma}, \ \xi_{\rho\sigma}]
G_{\mu\rho, \nu\lambda\lambda}
\nonumber
\\
&&
\hspace*{15mm}
+(b_3-b_{12})
[G_{\mu\nu, \rho}, \ \xi_{\sigma\sigma}]
G_{\mu\lambda, \nu\rho\lambda}
+2(b_3-b_{12})
[G_{\mu\nu, \sigma}, \ \xi_{\rho\sigma}]
G_{\mu\lambda, \nu\rho\lambda}
\nonumber
\\
&&
\hspace*{15mm}
+(b_3-4b_{11})
[G_{\mu\lambda, \nu}, \ \xi_{\rho\lambda}]
G_{\mu\nu, \rho\sigma\sigma}
+(b_3-b_{12})
[G_{\mu\lambda, \rho}, \ \xi_{\nu\lambda}]
G_{\mu\nu, \rho\sigma\sigma}
\nonumber
\\
&&
\hspace*{15mm}
+(b_3-4b_{13})
[G_{\mu\lambda, \lambda}, \ \xi_{\nu\rho}]
G_{\mu\nu, \rho\sigma\sigma}
+2(b_4-2b_{13})
[G_{\mu\nu, \nu}, \ \xi_{\rho\sigma}]
G_{\mu\lambda, \rho\sigma\lambda}
\nonumber
\\
&&
\hspace*{15mm}
+2(2b_4-b_{12})
[G_{\mu\nu, \rho}, \ \xi_{\nu\sigma}]
G_{\mu\lambda, \rho\sigma\lambda}
+2(b_5-b_{13})
[G_{\mu\nu, \nu}, \ \xi_{\rho\rho}]
G_{\mu\sigma, \sigma\lambda\lambda}
\nonumber
\\
&&
\hspace*{15mm}
+(4b_5-b_{12})
[G_{\mu\nu, \rho}, \ \xi_{\nu\rho}]
G_{\mu\sigma, \sigma\lambda\lambda}
+(b_6-4b_{14})
[G_{\mu\nu}, \ \xi_{\rho\sigma}]
G_{\mu\rho, \nu\sigma\lambda\lambda}
\nonumber
\\
&&
\hspace*{15mm}
+2(b_6-b_7)
[G_{\mu\rho, \sigma\lambda}, \ \xi_{\nu\lambda}]
G_{\mu\nu, \rho\sigma}
+(b_6-2b_8)
[G_{\mu\rho, \lambda\lambda}, \ \xi_{\nu\sigma}]
G_{\mu\nu, \rho\sigma}
\nonumber
\\
&&
\hspace*{15mm}
+(b_7-4b_{14})
[G_{\mu\nu}, \ \xi_{\rho\sigma}]
G_{\mu\lambda, \nu\rho\sigma\lambda}
+(b_7-4b_9)
[G_{\mu\lambda, \nu\lambda}, \ \xi_{\rho\sigma}]
G_{\mu\nu, \rho\sigma}
\nonumber
\\
&&
\hspace*{15mm}
+2(b_7-b_{10})
[G_{\mu\lambda, \rho\lambda}, \ \xi_{\nu\sigma}]
G_{\mu\nu, \rho\sigma}
+(b_8-2b_{14})
[G_{\mu\nu}, \ \xi_{\rho\rho}]
G_{\mu\sigma, \nu\sigma\lambda\lambda}
\nonumber
\\
&&
\hspace*{15mm}
+(b_8-2b_9)
[G_{\mu\sigma, \nu\sigma}, \ \xi_{\lambda\lambda}]
G_{\mu\nu, \rho\rho}
+(2b_8-b_{10})
[G_{\mu\sigma, \sigma\lambda}, \ \xi_{\nu\lambda}]
G_{\mu\nu, \rho\rho}
\nonumber
\\
&&
\hspace*{15mm}
+(b_9-b_{14})
[G_{\mu\nu}, \ \xi_{\nu\rho}]
G_{\mu\rho, \lambda\lambda\sigma\sigma}
+(b_{10}-4b_{14})
[G_{\mu\nu}, \ \xi_{\nu\rho}]
G_{\mu\lambda, \rho\lambda\sigma\sigma}
\nonumber
\\
&&
\hspace*{15mm}
+2(b_1-2b_{14})
[G_{\mu\nu}, \ \xi_{\rho\sigma\lambda}]
G_{\mu\rho, \nu\sigma\lambda}
+2(b_2-b_{14})
[G_{\mu\nu}, \ \xi_{\rho\sigma\sigma}]
G_{\mu\rho, \nu\lambda\lambda}
\nonumber
\\
&&
\hspace*{15mm}
+(b_3-4b_{14})
[G_{\mu\nu}, \ \xi_{\rho\sigma\sigma}]
G_{\mu\lambda, \nu\rho\lambda}
+(b_3-4b_{14})
[G_{\mu\lambda}, \ \xi_{\nu\rho\lambda}]
G_{\mu\nu, \rho\sigma\sigma}
\nonumber
\\
&&
\hspace*{15mm}
+2(b_4-2b_{14})
[G_{\mu\nu}, \ \xi_{\nu\rho\sigma}]
G_{\mu\lambda, \rho\sigma\lambda}
+2(b_5-b_{14})
[G_{\mu\nu}, \ \xi_{\nu\rho\rho}]
G_{\mu\sigma, \sigma\lambda\lambda}
\nonumber
\\
&&
\hspace*{15mm}
+(b_6-4b_{11})
[G_{\mu\rho, \lambda}, \ \xi_{\sigma\lambda\lambda}]
G_{\mu\nu, \rho\sigma}
+(b_6-b_{12})
[G_{\mu\rho, \sigma}, \ \xi_{\nu\lambda\lambda}]
G_{\mu\nu, \rho\sigma}
\nonumber
\\
&&
\hspace*{15mm}
+2(b_6-b_{12})
[G_{\mu\rho, \lambda}, \ \xi_{\nu\sigma\lambda}]
G_{\mu\nu, \rho\sigma}
+(b_7-4b_{11})
[G_{\mu\lambda, \nu}, \ \xi_{\rho\sigma\lambda}]
G_{\mu\nu, \rho\sigma}
\nonumber
\\
&&
\hspace*{15mm}
+2(b_7-b_{12})
[G_{\mu\lambda, \rho}, \ \xi_{\nu\lambda\sigma}]
G_{\mu\nu, \rho\sigma}
+(b_7-4b_{13})
[G_{\mu\lambda, \lambda}, \ \xi_{\nu\rho\sigma}]
G_{\mu\nu, \rho\sigma}
\nonumber
\\
&&
\hspace*{15mm}
+(b_8-2b_{11})
[G_{\mu\sigma, \nu}, \ \xi_{\sigma\lambda\lambda}]
G_{\mu\nu, \rho\rho}
+(b_8-2b_{13})
[G_{\mu\sigma, \sigma}, \ \xi_{\nu\lambda\lambda}]
G_{\mu\nu, \rho\rho}
\nonumber
\\
&&
\hspace*{15mm}
+(2b_8-b_{12})
[G_{\mu\sigma, \lambda}, \ \xi_{\nu\sigma\lambda}]
G_{\mu\nu, \rho\rho}
+(4b_9-b_{12})
[G_{\mu\rho, \lambda}, \ \xi_{\lambda\sigma\sigma}]
G_{\mu\nu, \nu\rho}
\nonumber
\\
&&
\hspace*{15mm}
+(b_{10}-b_{12})
[G_{\mu\lambda, \rho}, \ \xi_{\lambda\sigma\sigma}]
G_{\mu\nu, \nu\rho}
+(b_{10}-4b_{13})
[G_{\mu\lambda, \lambda}, \ \xi_{\rho\sigma\sigma}]
G_{\mu\nu, \nu\rho}
\nonumber
\\
&&
\hspace*{15mm}
+2(b_{10}-b_{12})
[G_{\mu\lambda, \sigma}, \ \xi_{\rho\lambda\sigma}]
G_{\mu\nu, \nu\rho}
+(b_6-4b_{14})
[G_{\mu\rho}, \ \xi_{\nu\sigma\lambda\lambda}]
G_{\mu\nu, \rho\sigma}
\nonumber
\\
&&
\hspace*{15mm}
+(b_7-4b_{14})
[G_{\mu\lambda}, \ \xi_{\nu\rho\sigma\lambda}]
G_{\mu\nu, \rho\sigma}
+(b_8-2b_{14})
[G_{\mu\sigma}, \ \xi_{\nu\sigma\lambda\lambda}]
G_{\mu\nu, \rho\rho}
\nonumber
\\
&&
\hspace*{15mm}
+(b_9-b_{14})
[G_{\mu\rho}, \ \xi_{\lambda\lambda\sigma\sigma}]
G_{\mu\nu, \nu\rho}
+(b_{10}-4b_{14})
[G_{\mu\lambda}, \ \xi_{\rho\lambda\sigma\sigma}]
G_{\mu\nu, \nu\rho}
\nonumber
\\
&&
\hspace*{15mm}
+(4b_{11}-b_{12})
[G_{\mu\rho, \lambda}, \ \xi_{\nu\lambda\sigma\sigma}]
G_{\mu\nu, \rho}
+(b_{12}-4b_{13})
[G_{\mu\lambda, \lambda}, \ \xi_{\nu\rho\sigma\sigma}]
G_{\mu\nu, \rho}
\nonumber
\\
&&
\hspace*{15mm}
+(b_{11}-b_{14})
[G_{\mu\rho}, \ \xi_{\nu\lambda\lambda\sigma\sigma}]
G_{\mu\nu, \rho}
+(b_{12}-4b_{14})
[G_{\mu\lambda}, \ \xi_{\nu\rho\lambda\sigma\sigma}]
G_{\mu\nu, \rho}
\nonumber
\\
&&
\hspace*{15mm}
+(b_{13}-b_{14})
[G_{\mu\rho}, \ \xi_{\rho\sigma\sigma\lambda\lambda}]
G_{\mu\nu, \nu}
\Bigg\}.
\nonumber
\end{eqnarray}
Then we can determine the numerical coefficients consistently as
%
\renewcommand{\arraystretch}{2.0}
%
$$
\begin{array}{rclcrclcrclcrclcrclcrclcrcl}
b_2
&=&
\displaystyle
\frac{1}{2}b_1,
&\ &
b_3
&=&
2b_1,
&\ &
b_4
&=&
b_1,
&\ &
b_5
&=&
\displaystyle
\frac{1}{2}b_1,
&\ &
b_6
&=&
2b_1,
&\ &
b_7
&=&
2b_1,
&\ &
b_8
&=&
b_1,
\\
b_9
&=&
\displaystyle
\frac{1}{2}b_1,
&\ &
b_{10}
&=&
2b_1,
&\ &
b_{11}
&=&
\displaystyle
\frac{1}{2}b_1,
&\ &
b_{12}
&=&
2b_1,
&\ &
b_{13}
&=&
\displaystyle
\frac{1}{2}b_1,
&\ &
b_{14}
&=&
\displaystyle
\frac{1}{2}b_1,
&\ &
&&
\end{array}
$$
%
\renewcommand{\arraystretch}{1.6}
%
and come to the following second invariant Lagrangian
\begin{eqnarray}
\CL_4'
&=&
b_1\tr
\bigg(
G_{\mu\nu, \rho\sigma\lambda}G_{\mu\rho, \nu\sigma\lambda}
+\frac{1}{2}
G_{\mu\nu, \rho\sigma\sigma}G_{\mu\rho, \nu\lambda\lambda}
+2
G_{\mu\nu, \rho\sigma\sigma}G_{\mu\lambda, \nu\rho\lambda}
+
G_{\mu\nu, \nu\rho\sigma}G_{\mu\lambda, \rho\sigma\lambda}
\nonumber
\\
&&
\hspace*{11mm}
+\frac{1}{2}
G_{\mu\nu, \nu\rho\rho}G_{\mu\sigma, \sigma\lambda\lambda}
+2
G_{\mu\nu, \rho\sigma}G_{\mu\rho, \nu\sigma\lambda\lambda}
+2
G_{\mu\nu, \rho\sigma}G_{\mu\lambda, \nu\rho\sigma\lambda}
+
G_{\mu\nu, \rho\rho}G_{\mu\sigma, \nu\sigma\lambda\lambda}
\nonumber
\\
&&
\hspace*{11mm}
+\frac{1}{2}
G_{\mu\nu, \nu\rho}G_{\mu\rho, \lambda\lambda\sigma\sigma}
+2
G_{\mu\nu, \nu\rho}G_{\mu\lambda, \rho\lambda\sigma\sigma}
+\frac{1}{2}
G_{\mu\nu, \rho}G_{\mu\rho, \nu\lambda\lambda\sigma\sigma}
+2
G_{\mu\nu, \rho}G_{\mu\lambda, \nu\rho\lambda\sigma\sigma}
\nonumber
\\
&&
\hspace*{11mm}
+\frac{1}{2}
G_{\mu\nu, \nu}G_{\mu\rho, \rho\sigma\sigma\lambda\lambda}
+\frac{1}{2}
G_{\mu\nu}G_{\mu\rho, \nu\rho\sigma\sigma\lambda\lambda}
\bigg).
\end{eqnarray}
As we demonstrated so far,
these Lagrangians contain no subinvariance,
since the equation for coefficients $a$ and  $b$ has a unique
solution presented above.
Thus these construction completely fixes all coefficients in front of all
Lorentz structures participated in  $\CL_4$ and $\CL'_4$
and also proves that only two invariants $\CL_4$ and $\CL'_4$ are available.
It seems that the above construction of all possible Lorentz invariant
structures is difficult to extend to higher rank tensor fields.

\section{Higher Rank Tensor Gauge Fields}
\setcounter{equation}{0}
\setcounter{footnote}{0}

The progress in the construction of gauge and Lorentz invariant forms
for higher rank tensor gauge fields have been achieved  in the
approach proposed in \cite{Savvidy:2005fi, Savvidy:2005ki}, where
instead of constructing all possible Lorentz structures as in \cite{Savvidy:2005zm}
it was suggested
first to build a general gauge invariant tensor density and then by contraction to build
Lorentz invariants. In this way two invariant forms $\CL_{s+1}$ and $\CL'_{s+1}$
have been constructed
in \cite{Savvidy:2005fi, Savvidy:2005ki}. Therefore it is plausible that for
higher rank tensor gauge fields there exist only two independent gauge invariant forms
$\CL_{s+1}$ and $\CL'_{s+1}$.

In the present section we shall estimate the number of independent Lorentz invariant
structures appearing in $\CL_{s+1}$ and $\CL'_{s+1}$ and shall present
a compact form of its coefficients.
It will be demonstrated that the total number of the Lorentz independent structures
in the general Lagrangian for the rank-$s$ tensors grows as $s^2$.

The gauge invariant tensor density constructed in
\cite{Savvidy:2005fi, Savvidy:2005ki}
by expansion over the vector variable $e_{\mu}$ \cite{Savvidy:dv} is:
%
%
%
\begin{eqnarray}
\Big({\cal L}_{s+1}\Big)_{\nu_1\nu_2, \lambda_1\cdots\lambda_{2s}}
&=&
\tr
\Bigg(
\sum_{i=0}^{s-1}\sum_{p_1<p_2<\cdots<p_i}\!\!\!
G_{\mu\nu_1,
\lambda_{p_1}\lambda_{p_2}\cdots\lambda_{p_i}}
G_{\mu\nu_2,
\lambda_1\cdots\accentset{\vee}{\lambda_{p_1}}\cdots
\accentset{\vee}{\lambda_{p_2}}\cdots
\accentset{\vee}{\lambda_{p_i}}\cdots\lambda_{2s}}
\nonumber \\
&&
\hspace*{9mm}
+\frac{1}{2}\!\sum_{p_1<p_2<\cdots<p_s}\!\!\!
G_{\mu\nu_1, \lambda_{p_1}\lambda_{p_2}\cdots\lambda_{p_s}}
G_{\mu\nu_2,
\lambda_1\cdots\accentset{\vee}{\lambda_{p_1}}\cdots
\accentset{\vee}\lambda_{p_2}\cdots
\accentset{\vee}{\lambda_{p_s}}\cdots\lambda_{2s}}
\nonumber
\\
&&
\hspace*{9mm}
+(\nu_1\leftrightarrow\nu_2)
\Bigg).
\label{invariant_tensor_1}
\end{eqnarray}
It should be noted that
these tensors are totally symmetric within the indices
$(\nu_1, \nu_2)$ and $(\lambda_1, \lambda_2, \cdots, \lambda_{2s})$,
whereas these have no symmetric structures between the
indices $\nu_i$ and $\lambda_j$.
Contraction of indices by metric tensors  allows to
construct two Lorentz invariant forms ${\cal L}_{s+1}$
and ${\cal L}_{s+1}'$ \cite{Savvidy:2005fi, Savvidy:2005ki}
\begin{subequations}
\begin{eqnarray}
{\cal L}_{s+1}
&=&
\frac{1}{2^s}
\Big({\cal L}_{s+1}\Big)_{\nu_1\nu_2, \lambda_1\cdots\lambda_{2s}}
\Big(
\eta_{\nu_1\nu_2}\eta_{\lambda_1\lambda_2}
\Big)
\eta_{\lambda_3\lambda_4}
\cdots\eta_{\lambda_{2s-1}\lambda_{2s}},
\label{general_lagrangian_0}\\
{\cal L}_{s+1}'
&=&
\frac{1}{2^{s-1}}
\Big({\cal L}_{s+1}\Big)_{\nu_1\nu_2, \lambda_1\cdots\lambda_{2s}}
\Big(\eta_{\nu_1\lambda_1}\eta_{\nu_2\lambda_2}
\Big)
\eta_{\lambda_3\lambda_4}
\cdots\eta_{\lambda_{2s-1}\lambda_{2s}}.
\label{general_lagrangian_prime_0}
\end{eqnarray}
\end{subequations}
Let us perform an explicit contraction of the indices in the equations
(\ref{general_lagrangian_0}) and (\ref{general_lagrangian_prime_0}).
First let us consider the Lagrangian $\CL_{s+1}$.
After contraction by $\eta_{\nu_1\nu_2}$ in
(\ref{general_lagrangian_0}),
it might be convenient to classify independent terms
by the number of contractions within the first field strength tensor
$G_{\mu\nu, \alpha_1\alpha_1\alpha_2\alpha_2\cdots\alpha_j\alpha_j
\beta_1\beta_2\cdots\beta_k}(x)$ ($j$-contractions, $k$-going out).
Through this procedure we can determine the combinatorial numbers
as the coefficients\footnote{
We denote the symbol $[m]$ as Gauss symbol.
},
\begin{eqnarray}
\CL_{s+1}
&=&
\frac{1}{2^{s-1}}\tr
\Bigg(
\sum_{i=0}^{s-1}\sum_{j=0}^{[\frac{i}{2}]}
a\Big(s, i, j\Big)
\nonumber
\\
&&
\hspace*{25mm}
\times
G_{\mu\nu, \alpha_1\alpha_1\alpha_2\alpha_2\cdots\alpha_j\alpha_j
\beta_1\beta_2\cdots\beta_{i-2j}}
G_{\mu\nu, \gamma_1\gamma_1\gamma_2\gamma_2\cdots
\gamma_{s-i+j}\gamma_{s-i+j}\beta_1\beta_2\cdots\beta_{i-2j}}
\nonumber
\\
&&
\hspace*{16mm}
+\frac{1}{2}\sum_{j=0}^{[\frac{s}{2}]}
a\Big(s, s, j\Big)
\nonumber
\\
&&
\hspace*{25mm}
\times
G_{\mu\nu, \alpha_1\alpha_1\alpha_2\alpha_2\cdots\alpha_j\alpha_j
\beta_1\beta_2\cdots\beta_{s-2j}}
G_{\mu\nu, \gamma_1\gamma_1\gamma_2\gamma_2\cdots
\gamma_{j}\gamma_{j}\beta_1\beta_2\cdots\beta_{s-2j}}
\Bigg),
\label{general_lagrangian}
\end{eqnarray}
where the numerical coefficients $a\Big(s, i, j\Big)$ are given
by\footnote{
We use a convention of the binomial coefficients
$\scriptsize\left(
\begin{array}{c}
m \\
n
\end{array}
\right)$
as
$\scriptsize\left(
\begin{array}{c}
m \\
n
\end{array}
\right)=0$ for
$m<n$ or $n<0$.
}
\begin{equation}
a\Big(s, i, j\Big)
=
2^{i-2j}
\left(
\begin{array}{c}
s-j \\
i-2j
\end{array}
\right)
\!
\left(
\begin{array}{c}
s \\
j
\end{array}
\right).
\end{equation}
From this formula (\ref{general_lagrangian}),
we can also compute the total number of the independent quadratic
terms in the Lagrangian $\CL_{s+1}$ as\footnote{
These results can be applied for the case $s=0$ as well.
}
\begin{eqnarray}
{\hbox{total number}}
&=&
\sum_{i=0}^{s-1}
\bigg(\Big[\frac{i}{2}\Big]+1\bigg)
+\bigg(\Big[\frac{s}{2}\Big]+1\bigg)
\nonumber
\\
&=&
\left\{
%
\renewcommand{\arraystretch}{2.0}
\begin{array}{lcl}
\displaystyle
\frac{s^2}{4}+s+1
&\qquad& ({\hbox{$s$: even}}), \\
\displaystyle
\frac{s^2}{4}+s+\frac{3}{4}
&\qquad& ({\hbox{$s$: odd}}).
\end{array}
\right.
\end{eqnarray}
%
\renewcommand{\arraystretch}{1.6}

\vspace*{-4mm}

Next, we turn to consider the Lagrangian $\CL_{s+1}'$.
After the summation
of $\Big(\eta_{\nu_1\lambda_1}\eta_{\nu_2\lambda_2}\Big)$
in (\ref{general_lagrangian_prime_0}),
we can classify the terms into four groups,
such as
\begin{equation}
\tr\Big(G_{\mu\nu, \cdots}G_{\mu\rho, \nu\rho\cdots}\Big), \quad
\tr\Big(G_{\mu\nu, \nu\cdots}G_{\mu\rho, \rho\cdots}\Big), \quad
\tr\Big(G_{\mu\nu, \rho\cdots}G_{\mu\rho, \nu\cdots}\Big), \quad
\tr\Big(G_{\mu\nu, \nu\rho\cdots}G_{\mu\rho, \cdots}\Big).
\label{classification}
\end{equation}
Then we proceed to contract between the remaining indices
in all above terms.
In this manipulation also
it might be convenient to classify the
independent terms by the number of contractions within the first
field strength
$G_{\mu\nu,
\alpha_1\alpha_1\alpha_2\alpha_2
\cdots\alpha_j\alpha_j\beta_1\beta_2\cdots\beta_k}(x)$
($j$-contractions, $k$-going out)
in each term.
It should be noted that
the first and the last ones in the classification
(\ref{classification}) could be the same
if the first and the second field strengths
in all traces
would have the same number of indices,
i.e.\ the case for rank-$(s+2)$ field strength tensors.
Taking this fact into account,
we obtain the following form with the combinatorial numbers
as the coefficients,
\begin{eqnarray}
\CL_{s+1}'
&=&
\frac{1}{2^{s-2}}
\tr\Bigg(
\sum_{i=0}^{s-1}\sum_{j=0}^{[\frac{i}{2}]}
a\Big(s-1, i, j\Big)
\nonumber
\\
&&
\hspace*{25mm}
\times
G_{\mu\nu, \alpha_1\alpha_1\alpha_2\alpha_2\cdots
\alpha_j\alpha_j\beta_1\beta_2\cdots\beta_{i-2j}}
G_{\mu\rho, \nu\rho\gamma_1\gamma_1\gamma_2\gamma_2\cdots
\gamma_{s-i+j-1}\gamma_{s-i+j-1}\beta_1\beta_2\cdots\beta_{i-2j}}
\nonumber
\\
&&
\hspace*{13mm}
+\sum_{i=1}^{s-1}\sum_{j=0}^{[\frac{i-1}{2}]}
a\Big(s-1, i-1, j\Big)
\nonumber
\\
&&
\hspace*{25mm}
\times
G_{\mu\nu, \nu\alpha_1\alpha_1\alpha_2\alpha_2\cdots
\alpha_j\alpha_j\beta_1\beta_2\cdots\beta_{i-2j-1}}
G_{\mu\rho, \rho\gamma_1\gamma_1\gamma_2\gamma_2\cdots
\gamma_{s-i+j}\gamma_{s-i+j}\beta_1\beta_2\cdots\beta_{i-2j-1}}
\nonumber
\\
&&
\hspace*{13mm}
+\sum_{i=1}^{s-1}\sum_{j=0}^{[\frac{i-1}{2}]}
a\Big(s-1, i-1, j\Big)
\nonumber
\\
&&
\hspace*{25mm}
\times
G_{\mu\nu, \rho\alpha_1\alpha_1\alpha_2\alpha_2\cdots
\alpha_j\alpha_j\beta_1\beta_2\cdots\beta_{i-2j-1}}
G_{\mu\rho, \nu\gamma_1\gamma_1\gamma_2\gamma_2\cdots
\gamma_{s-i+j}\gamma_{s-i+j}\beta_1\beta_2\cdots\beta_{i-2j-1}}
\nonumber
\\
&&
\hspace*{13mm}
+\sum_{i=2}^{s-1}
\sum_{j=0}^{[\frac{i-2}{2}]}
a\Big(s-1, i-2, j\Big)
\nonumber
\\
&&
\hspace*{25mm}
\times G_{\mu\nu, \nu\rho\alpha_1\alpha_1\alpha_2\alpha_2\cdots
\alpha_j\alpha_j\beta_1\beta_2\cdots\beta_{i-2j-2}}
G_{\mu\rho, \gamma_1\gamma_1\gamma_2\gamma_2\cdots
\gamma_{s-i+j+1}\gamma_{s-i+j+1}\beta_1\beta_2\cdots\beta_{i-2j-2}}
\nonumber
\\
&&
\hspace*{13mm}
+\sum_{j=1}^{[\frac{s}{2}]}
a\Big(s-1, s, j\Big)
\nonumber
\\
&&
\hspace*{25mm}
\times
G_{\mu\nu, \alpha_1\alpha_1\alpha_2\alpha_2\cdots
\alpha_j\alpha_j\beta_1\beta_2\cdots\beta_{s-2j}}
G_{\mu\rho, \nu\rho\gamma_1\gamma_1\gamma_2\gamma_2\cdots
\gamma_{j-1}\gamma_{j-1}\beta_1\beta_2\cdots\beta_{s-2j}}
\nonumber
\\
&&
\hspace*{13mm}
+\frac{1}{2}\sum_{j=0}^{[\frac{s-1}{2}]}
a\Big(s-1, s-1, j\Big)
\nonumber
\\
&&
\hspace*{25mm}
\times
G_{\mu\nu, \nu\alpha_1\alpha_1\alpha_2\alpha_2\cdots
\alpha_j\alpha_j\beta_1\beta_2\cdots\beta_{s-2j-1}}
G_{\mu\rho, \rho\gamma_1\gamma_1\gamma_2\gamma_2\cdots
\gamma_{j}\gamma_{j}\beta_1\beta_2\cdots\beta_{s-2j-1}}
\nonumber
\\
&&
\hspace*{13mm}
+\frac{1}{2}\sum_{j=0}^{[\frac{s-1}{2}]}
a\Big(s-1, s-1, j\Big)
\nonumber
\\
&&
\hspace*{25mm}
\times
G_{\mu\nu, \rho\alpha_1\alpha_1\alpha_2\alpha_2\cdots
\alpha_j\alpha_j\beta_1\beta_2\cdots\beta_{s-2j-1}}
G_{\mu\rho, \nu\gamma_1\gamma_1\gamma_2\gamma_2\cdots
\gamma_{j}\gamma_{j}\beta_1\beta_2\cdots\beta_{s-2j-1}}
\Bigg).
\label{general_lagrangian_prime}
\end{eqnarray}
From this formula (\ref{general_lagrangian_prime}),
we can see the total number of the independent quadratic
terms in the Lagrangian $\CL_{s+1}'$ as\footnote{
These results can be applied for the cases $s=0$, $1$ and $2$ as well.
}
\begin{eqnarray}
{\hbox{total number}}
&=&
\sum_{i=0}^{s-1}
\bigg(\Big[\frac{i}{2}\Big]+1\bigg)
+\sum_{i=1}^{s-1}
\bigg(\Big[\frac{i-1}{2}\Big]+1\bigg)
+\sum_{i=1}^{s-1}
\bigg(\Big[\frac{i-1}{2}\Big]+1\bigg)
+\sum_{i=2}^{s-1}
\bigg(\Big[\frac{i-2}{2}\Big]+1\bigg)
\nonumber
\\
&&
+\Big[\frac{s}{2}\Big]
+\bigg(\Big[\frac{s-1}{2}\Big]+1\bigg)
+\bigg(\Big[\frac{s-1}{2}\Big]+1\bigg)
\nonumber
\\
&=&
\left\{
%
\renewcommand{\arraystretch}{2.0}
\begin{array}{lcl}
\displaystyle
s^2+\frac{3}{2}s
&\qquad& ({\hbox{$s$: even}}), \\
\displaystyle
s^2+\frac{3}{2}s+\frac{1}{2}
&\qquad& ({\hbox{$s$: odd}}).
\end{array}
\right.
\end{eqnarray}
%
\renewcommand{\arraystretch}{1.6}

\vspace*{-4mm}
\noindent
In both Lagrangians $\CL_{s+1}$ and $\CL_{s+1}'$
the total number of terms grows as $s^2$.

\section{Appendix}
\setcounter{equation}{0}
\setcounter{footnote}{0}

Let us get convinced that for $s=1,2$ and $3$ the
invariant tensor density (\ref{invariant_tensor_1})
and the formulas (\ref{general_lagrangian}) and (\ref{general_lagrangian_prime})
reproduce the lower rank Lagrangians:
\begin{itemize}
%
%
%
%
%
%

\item $s=1$ \
$\Big($ for second rank tensor gauge filed $\Big)$

invariant tensor:

\vspace*{-5mm}

\begin{equation}
\Big({\cal L}_2\Big)_{\nu_1\nu_2, \lambda_1\lambda_2}
=
\tr
\bigg(
G_{\mu\nu_1}G_{\mu\nu_2, \lambda_1\lambda_2}
+G_{\mu\nu_1, \lambda_1}G_{\mu\nu_2, \lambda_2}
+(\nu_1\leftrightarrow\nu_2)
\bigg).
\end{equation}

Lagrangians:

\vspace*{-10mm}

\begin{subequations}
\begin{eqnarray}
{\cal L}_2
&=&
\frac{1}{2}\Big({\cal L}_2\Big)_{\nu_1\nu_2, \lambda_1\lambda_2}
\eta_{\nu_1\nu_2}\eta_{\lambda_1\lambda_2}
\nonumber
\\
&=&
\tr\bigg(
G_{\mu\nu}G_{\mu\nu, \rho\rho}
+G_{\mu\nu, \rho}G_{\mu\nu, \rho}
\bigg),
\\
&&
\nonumber
\\
{\cal L}_2'
&=&
\Big({\cal L}_2\Big)_{\nu_1\nu_2, \lambda_1\lambda_2}
\eta_{\nu_1\lambda_1}\eta_{\nu_2\lambda_2}
\nonumber
\\
&=&
2\tr
\bigg(
G_{\mu\nu}
G_{\mu\rho, \nu\rho}
+\frac{1}{2}
G_{\mu\nu, \nu}
G_{\mu\rho, \rho}
+\frac{1}{2}
G_{\mu\nu, \rho}
G_{\mu\rho, \nu}
\bigg).
\end{eqnarray}
\end{subequations}

\vspace*{-6mm}

\item $s=2$ \
$\Big($ for third rank tensor gauge filed $\Big)$

invariant tensor:

\vspace*{-10mm}

\begin{eqnarray}
\Big({\cal L}_3\Big)_{\nu_1\nu_2, \lambda_1\lambda_1\lambda_3\lambda_4}
&=&
\tr
\bigg(
G_{\mu\nu_1}
G_{\mu\nu_2, \lambda_1\lambda_2\lambda_3\lambda_4}
\nonumber
\\
&&
\hspace*{7mm}
+G_{\mu\nu_1, \lambda_1}
G_{\mu\nu_2, \lambda_2\lambda_3\lambda_4}
+G_{\mu\nu_1, \lambda_2}
G_{\mu\nu_2, \lambda_1\lambda_3\lambda_4}
\nonumber
\\
&&
\hspace*{7mm}
+G_{\mu\nu_1, \lambda_3}
G_{\mu\nu_2, \lambda_1\lambda_2\lambda_4}
+G_{\mu\nu_1, \lambda_4}
G_{\mu\nu_2, \lambda_1\lambda_2\lambda_3}
\nonumber
\\
&&
\hspace*{7mm}
+G_{\mu\nu_1, \lambda_1\lambda_2}
G_{\mu\nu_2, \lambda_3\lambda_4}
+G_{\mu\nu_1, \lambda_1\lambda_3}
G_{\mu\nu_2, \lambda_2\lambda_4}
\nonumber
\\
&&
\hspace*{7mm}
+G_{\mu\nu_1, \lambda_1\lambda_4}
G_{\mu\nu_2, \lambda_2\lambda_3}
\nonumber
\\
&&
\hspace*{7mm}
+(\nu_1\leftrightarrow\nu_2)
\bigg).
\end{eqnarray}

Lagrangians:

\vspace*{-10mm}

\begin{subequations}
\begin{eqnarray}
{\cal L}_3
&=&
\frac{1}{4}
\Big({\cal L}_3\Big)_{\nu_1\nu_2, \lambda_1\lambda_2\lambda_3\lambda4}
\eta_{\nu_1\nu_2}\eta_{\lambda_1\lambda_2}\eta_{\lambda_3\lambda_4}
\nonumber
\\
&=&
\frac{1}{2}\tr
\bigg(
G_{\mu\nu}G_{\mu\nu, \rho\rho\sigma\sigma}
\nonumber
\\
&&
\hspace*{10mm}
+4G_{\mu\nu, \rho}G_{\mu\nu ,\sigma\sigma\rho}
\nonumber
\\
&&
\hspace*{10mm}
+2G_{\mu\nu, \rho\sigma}G_{\mu\nu, \rho\sigma}
+G_{\mu\nu, \rho\rho}G_{\mu\nu, \sigma\sigma}
\bigg),
\\
&&
\nonumber
\\
{\cal L}_3'
&=&
\frac{1}{2}
\Big({\cal L}_3\Big)_{\nu_1\nu_2, \lambda_1\lambda_2\lambda_3\lambda_4}
\eta_{\nu_1\lambda_1}\eta_{\nu_2\lambda_2}\eta_{\lambda_3\lambda_4}
\nonumber
\\
&=&
\tr
\bigg(
G_{\mu\nu}G_{\mu\rho, \nu\rho\sigma\sigma}
\nonumber
\\
&&
\hspace*{7mm}
+2G_{\mu\nu, \sigma}G_{\mu\rho, \nu\rho\sigma}
+G_{\mu\nu, \nu}G_{\mu\rho, \rho\sigma\sigma}
+G_{\mu\nu, \rho}G_{\mu\rho, \nu\sigma\sigma}
\nonumber
\\
&&
\hspace*{7mm}
+G_{\mu\nu, \sigma\sigma}G_{\mu\rho, \nu\rho}
+G_{\mu\nu, \nu\sigma}G_{\mu\rho, \rho\sigma}
+G_{\mu\nu, \rho\sigma}G_{\mu\rho, \nu\sigma}
\bigg).
\end{eqnarray}
\end{subequations}

\vspace*{-6mm}

\item $s=3$ \
$\Big($ for fourth rank tensor gauge filed $\Big)$

invariant tensor:
\begin{eqnarray}
\Big({\cal L}_4\Big)
_{\nu_1\nu_2, \lambda_1\lambda_2\lambda_3\lambda_4\lambda_5\lambda_6}
&=&
\tr
\bigg(
G_{\mu\nu_1}
G_{\mu\nu_2, \lambda_1\lambda_2\lambda_3\lambda_4\lambda_5\lambda_6}
\nonumber
\\
&&
\hspace*{7.5mm}
+G_{\mu\nu_1, \lambda_1}
G_{\mu\nu_2, \lambda_2\lambda_3\lambda_4\lambda_5\lambda_6}
+G_{\mu\nu_1, \lambda_2}
G_{\mu\nu_2, \lambda_1\lambda_3\lambda_4\lambda_5\lambda_6}
\nonumber
\\
&&
\hspace*{7.5mm}
+G_{\mu\nu_1, \lambda_3}
G_{\mu\nu_2, \lambda_1\lambda_2\lambda_4\lambda_5\lambda_6}
+G_{\mu\nu_1, \lambda_4}
G_{\mu\nu_2, \lambda_1\lambda_2\lambda_3\lambda_5\lambda_6}
\nonumber
\\
&&
\hspace*{7.5mm}
+G_{\mu\nu_1, \lambda_5}
G_{\mu\nu_2, \lambda_1\lambda_2\lambda_3\lambda_4\lambda_6}
+G_{\mu\nu_1, \lambda_6}
G_{\mu\nu_2, \lambda_1\lambda_2\lambda_3\lambda_4\lambda_5}
\nonumber
\\
&&
\hspace*{7.5mm}
+G_{\mu\nu_1, \lambda_1\lambda_2}
G_{\mu\nu_2, \lambda_3\lambda_4\lambda_5\lambda_6}
+G_{\mu\nu_1, \lambda_1\lambda_3}
G_{\mu\nu_2, \lambda_2\lambda_4\lambda_5\lambda_6}
\nonumber
\\
&&
\hspace*{7.5mm}
+G_{\mu\nu_1, \lambda_1\lambda_4}
G_{\mu\nu_2, \lambda_2\lambda_3\lambda_5\lambda_6}
+G_{\mu\nu_1, \lambda_1\lambda_5}
G_{\mu\nu_2, \lambda_2\lambda_3\lambda_4\lambda_6}
\nonumber
\\
&&
\hspace*{7.5mm}
+G_{\mu\nu_1, \lambda_1\lambda_6}
G_{\mu\nu_2, \lambda_2\lambda_3\lambda_4\lambda_5}
+G_{\mu\nu_1, \lambda_2\lambda_3}
G_{\mu\nu_2, \lambda_1\lambda_4\lambda_5\lambda_6}
\nonumber
\\
&&
\hspace*{7.5mm}
+G_{\mu\nu_1, \lambda_2\lambda_4}
G_{\mu\nu_2, \lambda_1\lambda_3\lambda_5\lambda_6}
+G_{\mu\nu_1, \lambda_2\lambda_5}
G_{\mu\nu_2, \lambda_1\lambda_3\lambda_4\lambda_6}
\nonumber
\\
&&
\hspace*{7.5mm}
+G_{\mu\nu_1, \lambda_2\lambda_6}
G_{\mu\nu_2, \lambda_1\lambda_3\lambda_4\lambda_5}
+G_{\mu\nu_1, \lambda_3\lambda_4}
G_{\mu\nu_2, \lambda_1\lambda_2\lambda_5\lambda_6}
\nonumber
\\
&&
\hspace*{7.5mm}
+G_{\mu\nu_1, \lambda_3\lambda_5}
G_{\mu\nu_2, \lambda_1\lambda_2\lambda_4\lambda_6}
+G_{\mu\nu_1, \lambda_3\lambda_6}
G_{\mu\nu_2, \lambda_1\lambda_2\lambda_4\lambda_5}
\nonumber
\\
&&
\hspace*{7.5mm}
+G_{\mu\nu_1, \lambda_4\lambda_5}
G_{\mu\nu_2, \lambda_1\lambda_2\lambda_3\lambda_6}
+G_{\mu\nu_1, \lambda_4\lambda_6}
G_{\mu\nu_2, \lambda_1\lambda_2\lambda_3\lambda_5}
\nonumber
\\
&&
\hspace*{7.5mm}
+G_{\mu\nu_1, \lambda_5\lambda_6}
G_{\mu\nu_2, \lambda_1\lambda_2\lambda_3\lambda_4}
\nonumber
\\
&&
\hspace*{7.5mm}
+G_{\mu\nu_1, \lambda_1\lambda_2\lambda_3}
G_{\mu\nu_2, \lambda_4\lambda_5\lambda_6}
+G_{\mu\nu_1, \lambda_1\lambda_2\lambda_4}
G_{\mu\nu_2, \lambda_3\lambda_5\lambda_6}
\nonumber
\\
&&
\hspace*{7.5mm}
+G_{\mu\nu_1, \lambda_1\lambda_2\lambda_5}
G_{\mu\nu_2, \lambda_3\lambda_4\lambda_6}
+G_{\mu\nu_1, \lambda_1\lambda_2\lambda_6}
G_{\mu\nu_2, \lambda_3\lambda_4\lambda_5}
\nonumber
\\
&&
\hspace*{7.5mm}
+G_{\mu\nu_1, \lambda_1\lambda_3\lambda_4}
G_{\mu\nu_2, \lambda_2\lambda_5\lambda_6}
+G_{\mu\nu_1, \lambda_1\lambda_3\lambda_5}
G_{\mu\nu_2, \lambda_2\lambda_4\lambda_6}
\nonumber
\\
&&
\hspace*{7.5mm}
+G_{\mu\nu_1, \lambda_1\lambda_3\lambda_6}
G_{\mu\nu_2, \lambda_2\lambda_4\lambda_5}
+G_{\mu\nu_1, \lambda_1\lambda_4\lambda_5}
G_{\mu\nu_2, \lambda_2\lambda_3\lambda_6}
\nonumber
\\
&&
\hspace*{7.5mm}
+G_{\mu\nu_1, \lambda_1\lambda_4\lambda_6}
G_{\mu\nu_2, \lambda_2\lambda_3\lambda_5}
+G_{\mu\nu_1, \lambda_1\lambda_5\lambda_6}
G_{\mu\nu_2, \lambda_2\lambda_3\lambda_4}
\nonumber
\\
&&
\hspace*{7.5mm}
+(\nu_1\leftrightarrow\nu_2)
\bigg).
\end{eqnarray}

Lagrangians:

\vspace*{-10mm}

\begin{subequations}
\begin{eqnarray}
{\cal L}_4
&=&
\frac{1}{8}
\Big({\cal L}_4\Big)_{\nu_1\nu_2,
\lambda_1\lambda_2\lambda_3\lambda_4\lambda_5\lambda_6}
\eta_{\nu_1\nu_2}\eta_{\lambda_1\lambda_2}\eta_{\lambda_3\lambda_4}
\eta_{\lambda_5\lambda_6}
\nonumber
\\
&=&
\frac{1}{4}
\tr
\bigg(
G_{\mu\nu}
G_{\mu\nu, \lambda\lambda\rho\rho\sigma\sigma}
+6
G_{\mu\nu, \lambda}
G_{\mu\nu, \rho\rho\sigma\sigma\lambda}
\nonumber
\\
&&
\hspace*{10mm}
+12
G_{\mu\nu, \lambda\rho}G_{\mu\nu, \sigma\sigma\lambda\rho}
+3
G_{\mu\nu, \lambda\lambda}G_{\mu\nu, \rho\rho\sigma\sigma}
\nonumber
\\
&&
\hspace*{10mm}
+4
G_{\mu\nu, \lambda\rho\sigma}G_{\mu\nu, \lambda\rho\sigma}
+6
G_{\mu\nu, \lambda\lambda\rho}G_{\mu\nu, \sigma\sigma\rho}
\bigg),
\\
&&
\nonumber
\\
{\cal L}_4'
&=&
\frac{1}{4}
\Big({\cal L}_4\Big)_{\nu_1\nu_2,
\lambda_1\lambda_2\lambda_3\lambda_4\lambda_5\lambda_6}
\eta_{\nu_1\lambda_1}\eta_{\nu_2\lambda_2}\eta_{\lambda_3\lambda_4}
\eta_{\lambda_5\lambda_6}
\nonumber
\\
&=&
\frac{1}{2}
\tr
\bigg(
G_{\mu\nu}
G_{\mu\rho, \nu\rho\lambda\lambda\sigma\sigma}
+4
G_{\mu\nu, \lambda}
G_{\mu\rho, \nu\rho\sigma\sigma\lambda}
+4
G_{\mu\nu, \lambda\sigma}
G_{\mu\rho, \nu\rho\lambda\sigma}
+2
G_{\mu\nu, \lambda\lambda}
G_{\mu\rho, \nu\rho\sigma\sigma}
\nonumber
\\
&&
\hspace*{10mm}
+
G_{\mu\nu, \nu}
G_{\mu\rho, \rho\lambda\lambda\sigma\sigma}
+4
G_{\mu\nu, \nu\lambda}
G_{\mu\rho, \rho\sigma\sigma\lambda}
\nonumber
\\
&&
\hspace*{10mm}
+
G_{\mu\nu, \rho}
G_{\mu\rho, \nu\lambda\lambda\sigma\sigma}
+4
G_{\mu\nu, \rho\lambda}
G_{\mu\rho, \nu\sigma\sigma\lambda}
\nonumber
\\
&&
\hspace*{10mm}
+
G_{\mu\nu, \nu\rho}
G_{\mu\rho, \lambda\lambda\sigma\sigma}
\nonumber
\\
&&
\hspace*{10mm}
+4
G_{\mu\nu, \sigma\sigma\lambda}
G_{\mu\rho, \nu\rho\lambda}
\nonumber
\\
&&
\hspace*{10mm}
+2
G_{\mu\nu, \nu\lambda\sigma}
G_{\mu\rho, \rho\lambda\sigma}
+
G_{\mu\nu, \nu\lambda\lambda}
G_{\mu\rho, \rho\sigma\sigma}
\nonumber
\\
&&
\hspace*{10mm}
+2
G_{\mu\nu, \rho\lambda\sigma}
G_{\mu\rho, \nu\lambda\sigma}
+
G_{\mu\nu, \rho\lambda\lambda}
G_{\mu\rho, \nu\sigma\sigma}
\bigg).
\end{eqnarray}
\end{subequations}
\end{itemize}
%

\noindent
{\Large{\bf Acknowledgments}}
\vspace{2mm}
We would like to thank T. Jonsson for interesting comments and discussions.
T.T. would like to thank the National Center for Scientific Research
``DEMOKRITOS'' for warm hospitality. His work is supported by
ENRAGE (European Network on Random Geometry), a Marie Curie
Research Training Network supported by the European Community's
Sixth Framework Programme, network contract MRTN-CT-2004-005616.


\end{document}